\documentclass[letter,bibyear]{aa} 
\usepackage{epsfig}
\usepackage{amsmath}
\usepackage{subfigure}
\usepackage{natbib}
\usepackage{multirow}
\usepackage{color}
\usepackage{longtable}
\usepackage{graphicx}
\usepackage{epstopdf}
\usepackage{booktabs}
\usepackage{txfonts}
\usepackage[utf8]{inputenc}%

\newcommand{\jmst}{J.~Mol.~Struct.}   

\begin{document}

\title{Space and laboratory observation of the deuterated cyanomethyl radical HDCCN \thanks{Based on observations carried out
with the Yebes 40 m telescope (projects 19A003,
20A014, and 20D15) and the Institut de Radioastronomie Millim\'etrique (IRAM) 30 m telescope. The 40 m
radiotelescope at Yebes Observatory is operated by the Spanish Geographic Institute
(IGN, Ministerio de Transportes, Movilidad y Agenda Urbana).}}
\titlerunning{HDCCN}
\authorrunning{Cabezas et al.}

\author{
C.~Cabezas\inst{1},
Y.~Endo\inst{2},
E. Roueff\inst{3},
N.~Marcelino\inst{1},
M.~Ag\'undez\inst{1},
B.~Tercero\inst{4,5}, and
J.~Cernicharo\inst{1}
}

\institute{Grupo de Astrof\'isica Molecular, Instituto de F\'isica Fundamental (IFF-CSIC), C/ Serrano 121, 28006 Madrid, Spain.
\email carlos.cabezas@csic.es; jose.cernicharo@csic.es
\and Department of Applied Chemistry, Science Building II, National Chiao Tung University, 1001 Ta-Hsueh Rd., Hsinchu 30010, Taiwan
\and LERMA, Observatoire de Paris, PSL Research University, CNRS, Sorbonne Universit\'es, 92190 Meudon, France
\and Observatorio Astron\'omico Nacional (IGN), C/ Alfonso XII, 3, 28014, Madrid, Spain.
\and Centro de Desarrollos Tecnol\'ogicos, Observatorio de Yebes (IGN), 19141 Yebes, Guadalajara, Spain.
}

\date{Received; accepted}

\abstract{Our observations of TMC-1 with the Yebes 40 m radio telescope in the 31.0-50.3 GHz range allowed us to detect a group of
unidentified lines, showing a complex line pattern indicative of an open-shell species. {}The observed frequencies of these lines and
the similarity of the spectral pattern with that of the 2$_{0,2}$-1$_{0,1}$ rotational transition of H$_2$CCN
indicate that the lines arise from the deuterated cyanomethyl radical, HDCCN. Using Fourier transform microwave spectroscopy
experiments combined with electric discharges, we succeeded in producing the radical HDCCN in the laboratory and observed
its 1$_{0,1}$-0$_{0,0}$ and 2$_{0,2}$-1$_{0,1}$ rotational transitions. From our observations and assuming a rotational temperature of 5 K,
we derive an abundance ratio H$_2$CCN/HDCCN=20$\pm$4. The high abundance of the deuterated form of H$_2$CCN is well accounted for by a
standard gas-phase model, in which deuteration is driven by deuteron transfer from the H$_2$D$^+$ molecular ion.}


\maketitle

\section{Introduction}

Recent developments of radio astronomical equipments allow for very sensitive line surveys of molecular sources,
where weak lines arising from low-abundance species can be detected. These surveys are the best tool to map out
the molecular makeup of astronomical sources by identifying new chemical species. Discovering spectral
features of new molecules initially requires a detailed analysis of the spectral contribution of the already
identified species, including their isotopologs and vibrationally excited states. In this manner, once
all the lines coming from known species are assigned, unidentified lines emerge and open the opportunity
to discover new molecules and to deepen in our understanding of the chemical complexity of molecular clouds.

Deuterated molecules show remarkably high D/H abundance ratios
\citep{Ceccarelli2007} even though the deuterium abundance in space is $1.5 \times 10^{-5}$ relative to hydrogen \citep{Linsky2003}.
Hence, deuterated species of abundant interstellar molecules can also contribute to the spectral richness of line surveys. This fact makes the astronomical identification of these deuterated species of utmost importance, not only to gain knowledge about unidentified lines in the lines surveys but also to understand how the deuterium fractionation works and ultimately to constrain molecular formation pathways.

Cyanomethyl radical, H$_2$CCN, has been detected in the cold dark cloud TMC-1 \citep{Saito1988,Irvine1988},
where its abundance is relatively high. Hence, the deuterated isotopologs of H$_2$CCN are good candidates to be
observed in this source. As has been shown previously, a new high sensitivity line survey on TMC-1 has allowed the identification of several
transient species such as the anions C$_3$N$^-$ and C$_5$N$^-$ \citep{Cernicharo2020a} and the protonated molecules HC$_5$NH$^+$\citep{Marcelino2020}, HC$_3$O$^+$ \citep{Cernicharo2020b}, CH$_3$CO$^+$ \citep{Cernicharo2020c}, and HC$_3$S$^+$ \citep{Cernicharo2020d}.

In this Letter, we report the identification of spectral lines of the deuterated species HDCCN in TMC-1
based on ab initio calculations and accurate line frequencies from laboratory experiments.
Column densities for H$_2$CCN and HDCCN species are derived
from our observations and the most updated chemical models are used to diagnose the deuterium
fractionation ratios for the most abundant molecules in TMC-1.

\section{Observations} \label{observations}

The observations of TMC-1 ($\alpha_{J2000}=4^{\rm h} 41^{\rm  m} 41.9^{\rm s}$ and $\delta_{J2000}=+25^\circ 41' 27.0''$)
on the Q band presented in this work were performed in several sessions between November 2019 and October 2020.
They were carried out using a set of new receivers, which were built within the Nanocosmos
project\footnote{\texttt{https://nanocosmos.iff.csic.es/}} and installed at the Yebes 40 m radio telescope (hereafter,
Yebes 40 m).

The Q band receiver consists of two HEMT cold amplifiers covering the
31.0-50.3 GHz band with horizontal and vertical polarizations. The temperatures of the receiver
vary from 22 K at 32 GHz to 42 K at 50 GHz. The spectrometers formed by $2\times8\times2.5$ GHz FFTs
provide a spectral resolution of 38.1 kHz and cover the whole Q band in both polarizations.

\begin{figure}[]
\includegraphics[scale=0.65]{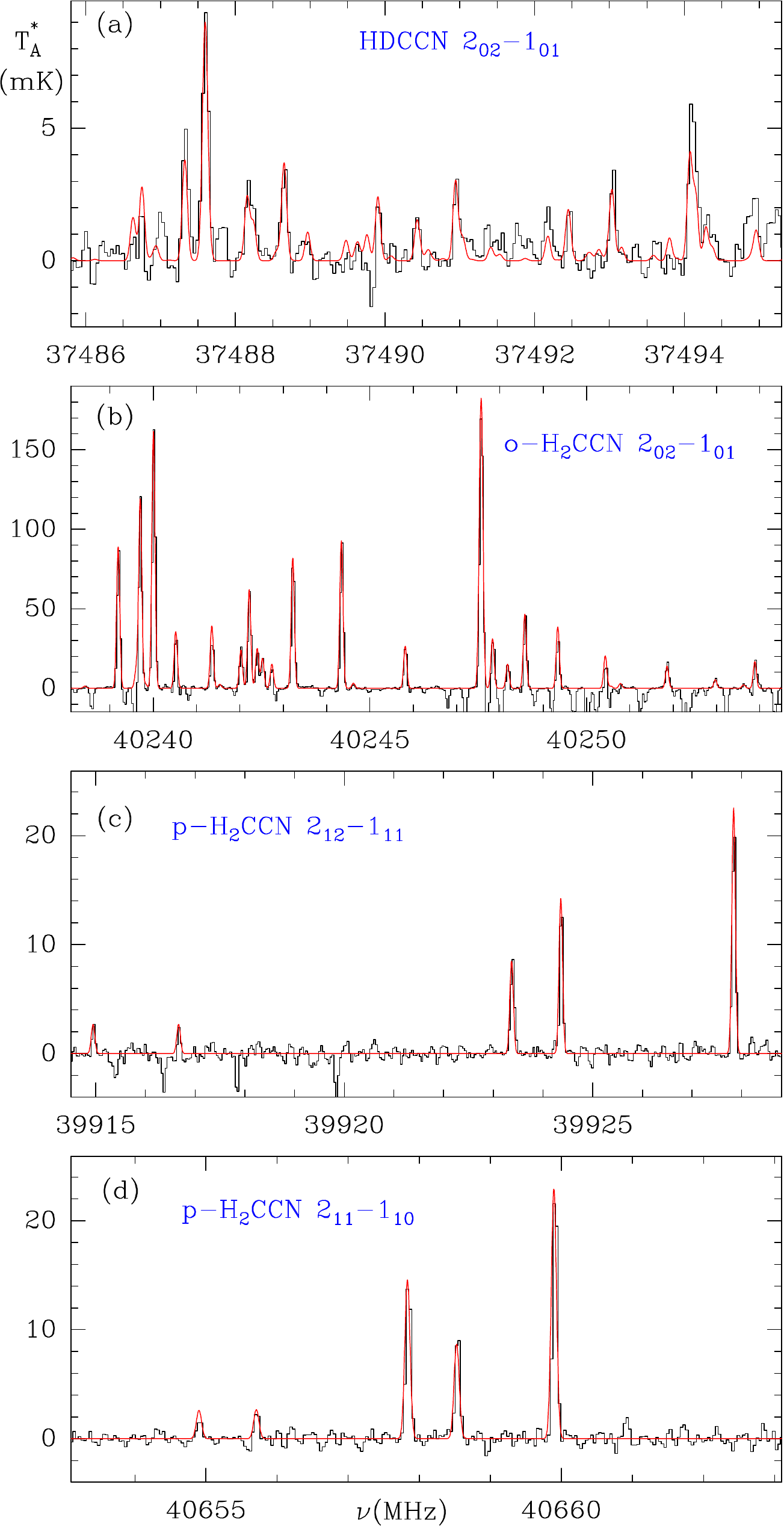}
\centering
\caption{ (a) Observed group of lines assigned to the hyperfine structure
of the 2$_{0,2}$-1$_{0,1}$ transition of HDCCN (black histogram). (b) The observed hyperfine structure of the same transition of o-H$_2$CCN. (c) Selected hyperfine
components of the 2$_{12}$-1$_{11}$ transition of p-H$_2$CCN. (d) Selected hyperfine
components of the 2$_{11}$-1$_{10}$ transition of p-H$_2$CCN. In all panels,
the abscissa represents the rest
frequency (in MHz) assuming a local standard of rest velocity for the source of
5.83 km\,s$^{-1}$ \citep{Cernicharo2020e}. The ordinate is the antenna temperature
in mK. The red curve in each panel represents
the best model fit to the emission of this species in TMC-1 for a rotational temperature
of 10 K. A similar fit is obtained for other rotational temperatures (see Sec.~\ref{CD_HDCCN} and Table~\ref{tab_CD}).
Negative
features are due to the folding of the frequency-switching data.}
\label{fig_hdccn}
\end{figure}

We observed different frequency coverages of  31.08-49.52 GHz and 31.98-50.42
GHz. This permitted us to check for spurious
ghosts produced in the down-conversion chain, in which the signal coming from the receiver was down-converted to 1-19.5 GHz and
then split into eight bands with a coverage of 2.5 GHz, each of which were analyzed by the FFTs.
Frequency switching with a frequency throw of 10\,MHz or 8\,MHz was chosen as the observing procedure.
The nominal spectral resolution of 38.1 kHz was left unchanged for the final spectra. The sensitivity, $\sigma$,
was derived by removing a polynomial baseline in velocity windows of 24 km\,s$^{-1}$ wide, centered on
each observed line, and varied along the
Q band between $\sim$0.4 mK (31 GHz), $\sim$1.0 mK (43 GHz), and $\sim$2.5 mK (50 GHz).

The intensity scale, antenna temperature
($T_A^*$) used in this work was calibrated using two absorbers at different temperatures and the
atmospheric transmission model ATM \citep{Cernicharo1985, Pardo2001}.
Calibration uncertainties were adopted to be 10\% based on the observed repeatability of the
line intensities between different observing runs.
All data were analyzed using the GILDAS package\footnote{\texttt{http://www.iram.fr/IRAMFR/GILDAS}}.

\section{Results} \label{sec:results}

\subsection{TMC-1 observations} \label{astro_results}

Among the unidentified lines of the line survey carried out by our group toward TMC-1 using the
Yebes 40 m radio telescope, we found a group of lines
with a spectral pattern reminiscent of that produced by a radical with a complex
hyperfine structure (see Fig.~\ref{fig_hdccn}a). The pattern is very similar
to the 2$_{0,2}$-1$_{0,1}$ rotational transition of H$_2$CCN
(see Fig.~\ref{fig_hdccn}b), which lies at a slightly higher frequency. These facts suggest that the group of
unidentified lines
could arise from the deuterated species of cyanomethyl radical. In order to check
this hypothesis we performed ab initio calculations (see Appendix~\ref{ab_initio}) for HDCCN using the
spectroscopic available information on H$_2$CCN \citep{Saito1997,Ozeki2004} to scale the spectroscopic parameters calculated.
The frequency of the 2$_{0,2}$-1$_{0,1}$ transition of HDCCN was predicted in the same frequency
region where the group of unidentified lines appear. The hyperfine structure of this transition of HDCCN is slightly
different from that of H$_2$CCN resulting from the replacement of a hydrogen nucleus by a deuteron. Motivated
by the good agreement between calculated and observed frequency and spectral pattern,
we performed additional ab initio calculations and laboratory experiments
to assign the complicated hyperfine structure of the rotational spectrum of HDCCN (see Sec.~\ref{labo} and Appendix~\ref{ab_initio}).
These experiments fully confirm that the carrier of the lines shown in Fig.~\ref{fig_hdccn}a is HDCCN.

\subsection{Laboratory measurements}
\label{labo}

\begin{figure}
\includegraphics[scale=0.40]{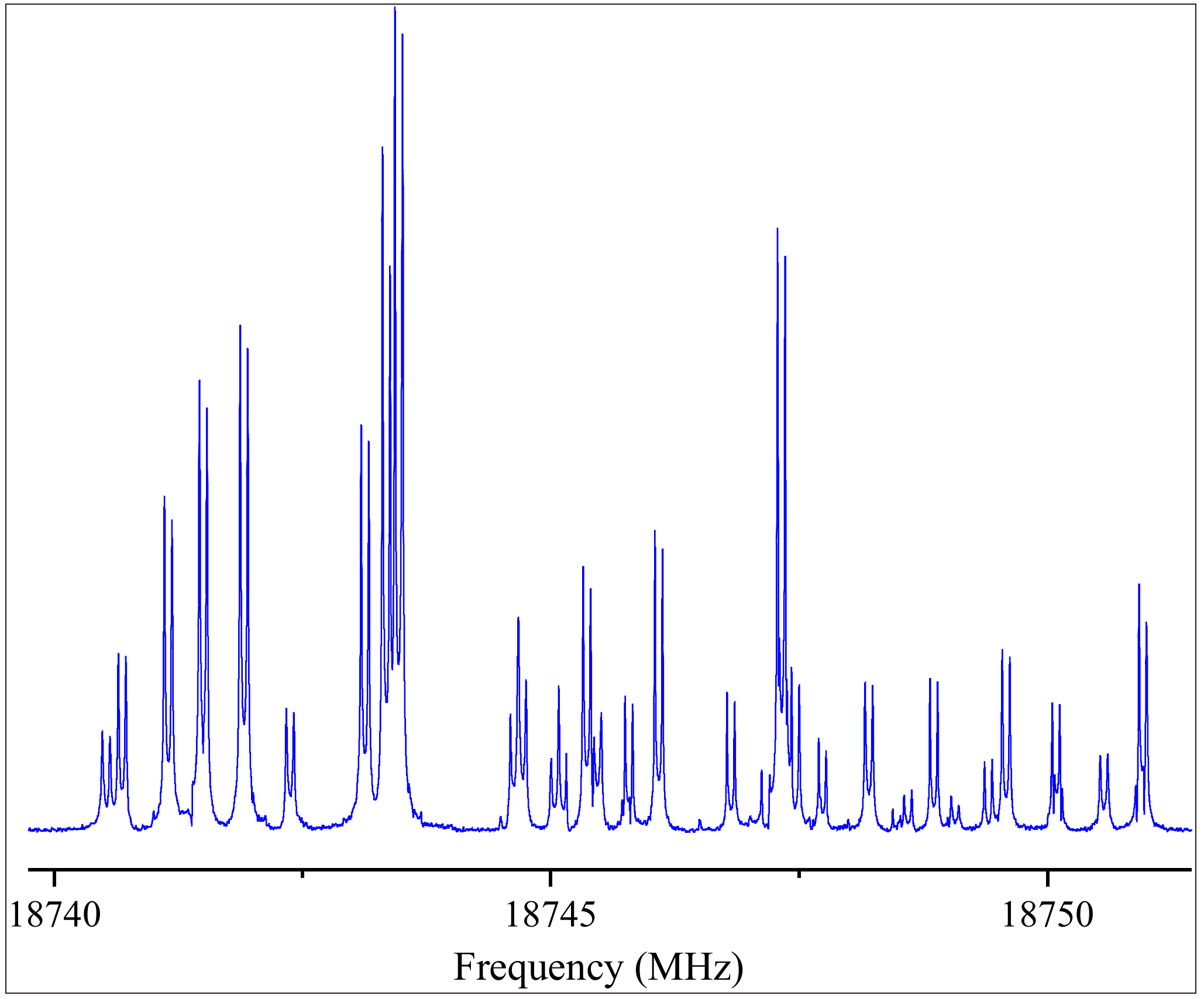}
\centering
\caption{ Section of 10 MHz of the FTMW spectrum of HDCCN showing 36 hyperfine components of the $N_{K_{a,}K_{c}}$= 1$_{0,1}$-0$_{0,0}$
pure rotational transition. The spectrum was
achieved by 100 shots of accumulation and a step scan of 0.5 MHz with a repetition rate of 10 Hz. The coaxial
arrangement of the adiabatic expansion and resonator axis produces an instrumental Doppler doubling.
The resonances frequencies are calculated as the average of the two Doppler components.}
\label{FTMW}
\end{figure}

The HDCCN radical was produced in a supersonic jet by a pulsed electric discharge of a gas mixture
of CH$_3$CN (0.2\%) and D$_2$ ( 5\%) diluted in Ar. The gas mixture was expanded into a Fabry-Perot
cavity of an Fourier transform microwave (FTMW) spectrometer \citep{Endo1994,Cabezas2016} with a
backing pressure at 2 atm.  Synchronized to the gas expansion, a pulse voltage of 1.0 kV with a duration
of 450 $\mu$s was applied
between stainless steel electrodes attached to the exit of the pulsed discharge nozzle (PDN).
For measurements of the paramagnetic lines, the Earth’s magnetic field was canceled using three sets of
Helmholtz coils placed
perpendicularly to one another.
Since the PDN is arranged parallel to the cavity of the spectrometer, it is possible to resolve small
hyperfine
splittings by suppressing
the Doppler broadening of spectral lines.

The $N_{K_{a,}K_{c}}$= 1$_{0,1}$-0$_{0,0}$ pure rotational transition was observed by FTMW spectroscopy while the
$N_{K_{a,}K_{c}}$= 2$_{0,2}$-1$_{0,1}$ transition was measured utilizing
a MW-MW double-resonance (DR) technique \citep{Sumiyoshi2005}. In DR measurements, the  signal of a
known rotational transition
is monitored with ordinary FTMW spectroscopy and when the pump radiation is resonant to a transition,
connecting the upper or lower
level of the monitored microwave transition, a DR spectrum is observed as a change of the
intensity of the monitored microwave signal.
First, a spectral scan of 100 MHz was done around the predicted frequency region for the $N_{K_{a,}K_{c}}$= 1$_{0,1}$-0$_{0,0}$ pure
rotational transition
to observe all the expected hyperfine components. The recorded spectrum for $N_{K_{a,}K_{c}}$= 1$_{0,1}$-0$_{0,0}$ transition
is shown in Figure \ref{FTMW}, where
a total of 66 hyperfine components were observed. Other 63 hyperfine components for
the $N_{K_{a,}K_{c}}$= 2$_{0,2}$-1$_{0,1}$ rotational transition
could be measured using the DR technique. Hence, a total of 129 hyperfine components (see Appendix \ref{lab_freq} and Table \ref{tab_lab_freq}) were analyzed
using a Hamiltonian with the following form:
\begin{equation}
H = H_{rot} + H_{sr} + H_{mhf} + H_Q,
\end{equation}
where $H_{rot}$ contains rotational and centrifugal distortion parameters, $H_{sr}$ is the spin-rotational term,
$H_{mhf}$ represents the magnetic hyperfine coupling interaction term due to the nitrogen, hydrogen, and
deuterium nuclei, and finally $H_Q$ represents the nuclear electric quadrupole interaction due to nitrogen
and deuterium nuclei. The coupling scheme used is $J$ = $N$ + $S$, $F_1$= $J$ + $I$(N), $F_2$= $F_1$ +
$I$(H) and $F$ = $F_2$ + $I$(D), and thus each energy level is denoted by seven quantum numbers; $N$,
$J$, $K_a$, $K_c$, $F_1$, $F_2$, and $F$.
The results obtained from the least-squares analysis are shown in Table \ref{spec_const}. We floated the
$(B+C)/2$ parameter, $\Delta_N$ centrifugal constant, $\varepsilon_{bb}$ spin-rotation coupling constant, $a_F$ and $T_{aa}$
magnetic hyperfine constants for nitrogen, hydrogen,
and deuterium nuclei, and the $\chi_{aa}$ nuclear electric quadrupole constant for nitrogen nucleus.\ The
$A$ and $(B-C)/2$ rotational parameters, $\varepsilon_{aa}$ and $\varepsilon_{cc}$ spin-rotation coupling constants,
and the $\chi_{aa}$ and $\chi_{bb}$ nuclear electric quadrupole constants for deuterium nucleus
were kept fixed to the theoretical values calculated in this work. The rest of parameters were fixed to
the values obtained by \citet{Saito1997} and \citet{Ozeki2004} for H$_2$CCN radical. The standard deviation of the
fit is 4.1 kHz.

\begin{table}
\begin{center}
\begin{tiny}
\caption[]{Spectroscopic parameters for HDCCN (all in MHz).}
{\label{spec_const}
\begin{tabular}{lc|lc|lc}
\hline
\hline
\multicolumn{2}{c}{Determined$^a$}&\multicolumn{2}{c}{Scaled$^b$}&\multicolumn{2}{c}{Fixed$^c$}\\
\hline
Param.  & Value & Param.  & Value & Param.  & Value\\
\hline
$(B+C)/2$          & ~9373.36479(35)$^d$ & $A$                & ~195819.2  &  $\Delta_{NK}$     &  ~    0.41626  \\
$\Delta_N$         & ~  0.004379(48)     & $(B-C)/2$          & ~230.594     &  $\Delta_K$        &  ~    13.6709  \\
$\varepsilon_{bb}$ & ~-22.6229(16)       & $\varepsilon_{aa}$ & ~-454.559  &  $\delta_J$        &  ~  0.0001611  \\
$a_F$(N)           & ~  9.4965(11)       & $\varepsilon_{cc}$ & ~-1.883    &  $\delta_K$        &  ~      0.263  \\
$T_{aa}$(N)        & ~-15.6405(28)       & $T_{bb}$(D)        & ~ 1.9721   &  $\Phi_{KN}$       &  ~   -0.000719 \\
$\chi_{aa}$(N)     & ~ -4.1906(30)       & $\chi_{aa}$(D)     & ~  -0.010  &  $\phi_{K}$        &  ~   -0.0024   \\
$a_F$(H)           & ~-59.8339(41)       & $\chi_{bb}$(D)     & ~   0.105  &  $^S\Delta_{KN}$   &  ~    0.0105   \\
$T_{aa}$(H)        & ~-17.7902(25)       &                    &            &  $^S\Delta_{K}$    &  ~   0.14      \\
$a_F$(D)           & ~ -9.1500(10)       &                    &            &  $T_{bb}$(N)       &  ~  -12.4519   \\
$T_{aa}$(D)        & ~ -2.1833(22)       &                    &            &  $C_{aa}$(N)       &  ~    0.0273   \\
                   &                     &                    &            &  $C_{bb}$(N)       &  ~  0.0029     \\
                   &                     &                    &            &  $\chi_{bb}$(N)    &  ~  1.8327     \\
                   &                     &                    &            &  $T_{bb}$(H)       &  ~  14.6       \\
\hline
\end{tabular}
}
\end{tiny}
\end{center}
\tablefoot{
\tablefoottext{a}{This work.}
\tablefoottext{b}{Fixed to the scaled values calculated in this work (see text).}
\tablefoottext{c}{Fixed at the values obtained by \cite{Saito1997} and \cite{Ozeki2004}.}
\tablefoottext{d}{Numbers in parentheses represent the derived uncertainty ($1\,\sigma$) of the parameter in units of the last digit.}
}
\end{table}

\section{Discussion}

\subsection{H$_2$CCN and HDCCN column densities} \label{CD_HDCCN}

The $N$=2-1 line of H$_2$CCN presents a very complex hyperfine structure in addition to the presence of two
spin symmetry species, ortho and para. Fig.~\ref{fig_hdccn}b shows the 2$_{02}$-1$_{01}$ ortho line; the
ortho species corresponds to even values of the quantum number $K_a$. Figs. \ref{fig_hdccn}c and \ref{fig_hdccn}d
show a selected range of hyperfine components of the $N_{K_{a,}K_{c}}$=2$_{1,2}$-1$_{1,1}$ and 2$_{1,1}$-1$_{1,0}$ para lines ($K_a$ odd).  Observed line intensities for the ortho species
agree very well with those reported by \citet{Kaifu2004}. No lines of the para species are reported
in that work, which probably results from the limited sensitivity in their spectra. However, some of their unidentified features agree in frequency with the
strongest components of this species (see also \citealt{Ozeki2004}). The detection of all hyperfine
components above 1.5 mK of the two para transitions within our line survey confirms the presence
of the para species in TMC-1.

Frequency predictions for the ortho and para species of H$_2$CCN are from the CDMS catalog \citep{Muller2005}, which were implemented in the MADEX code \citep{Cernicharo2012} to compute column densities for these species.
Both species have to be considered separately as there are no radiative or collisional
transitions between them. The lowest energy level of the para species (1$_{1,1}$) is 14.1 K above the ortho ground level (0$_{00}$).

In our line survey we only observed one rotational transition of the ortho H$_2$CCN (2$_{0,2}$-1$_{0,1}$) and two transitions for the para species
(2$_{1,2}$-1$_{1,1}$ and 2$_{1,1}$-1$_{1,0}$), which involve levels with similar energies. Taking into account that collisional
rates are not available for this molecule, column
density determinations rely on the assumption of a rotational temperature, T$_r$, for the two species.
Similar assumptions have been adopted in previous studies of this molecule \citep{Irvine1988,
Vastel2015}. In TMC-1, the rotational temperature found for different molecules
varies between 5 K and 10 K (see, e.g., \citealt{Gratier2016,Cernicharo2020d,Cernicharo2020e,Marcelino2020},
and references therein). In particular, for H$_2$CCN, \cite{Gratier2016} find a rotational temperature
between 3 K and 5 K. We consider that the lowest value for T$_r$ implies a large opacity
for the $N$=2-1 line of the ortho species (see Fig. \ref{fig_hdccn}).
The ortho species of the molecule behaves, for low rotational temperatures, as
a linear rotor because only levels with $K_a$=0 are populated in cold clouds. The first $K_a$=2 level (2$_{20}$) is at
55.6 K above the ground level (0$_{0,0}$).
The situation is similar for the para species because only levels with $K_a$=1 are populated. The first $K_a$=3 level (3$_{3,0}$) is 110.3 K above the ground level (1$_{1,1}$).
Hence, the rotational partition function, Q$_r$, for both symmetry species has a nearly
linear dependence on the rotational temperature (T$_r$$\leq$30 K).

We computed the ortho and para column densities that best fit the observed transitions shown in Fig. \ref{fig_hdccn}b, \ref{fig_hdccn}c, and \ref{fig_hdccn}d and for
rotational temperatures between 5 K and 10 K.
In these figures the red curve represents the synthetic model spectrum computed at T$_r$=10 K.
The derived total column density of H$_2$CCN shows a very smooth variation between 5 K and  10 K. This
is due to the near linear behavior of the partition function with T$_r$ and the low energy
of the involved levels.
Therefore the synthetic model spectrum is practically the same for these rotational temperatures.
Adopting a value for T$_r$=5 K (see Appendix C), we derive $N$(H$_2$CCN)=(1.48$\pm$0.15)$\times$10$^{13}$ cm$^{-2}$ and
an ortho/para ratio of 7.7, which is much larger than the statistical ratio of 3. It seems, hence, that
the para species, which is higher in energy than the ortho, has suffered interconversion processes
through the interaction with XH$^+$ species such as H$_3$$^+$, HCO$^+$, or H$_3$O$^+$. Our derived
column density is in good agreement with that derived by \citet{Ozeki2004} and \citet{Gratier2016}.

The rotational partition
function shows a dependence on T$_r$$^{3/2}$ for HDCCN, and hence the column density shows
a larger dependence on the assumed T$_r$. Table \ref{tab_CD} gives the derived
column densities of $o$-H$_2$CCN, $p$-H$_2$CCN, and HDCCN for rotational temperatures between 5 K and 10 K. We derive $N$(HDCCN)=(7.4$\pm$0.7)$\times$10$^{11}$ cm$^{-2}$ and $N$(H$_2$CCN)/$N$(HDCCN)=20$\pm$4 for T$_r$=5 K.
For T$_r$=10 K, these values are (10.5$\pm$1.1)$\times$10$^{11}$ cm$^{-2}$ and 14$\pm$3, respectively.
The derived XH/XD abundance ratios for H$_2$CCN and other species are given in Table \ref{tab_H/D}
(see Appendix \ref{deuteration} and Table \ref{tab_deuterated}).

The H$_2$CCN ortho/para ratio was also derived in L1544 by \cite{Vastel2015}. However, the derived
values for the column density of the ortho and para species and their abundance ratio
show a strong dependency on the assumed rotational temperature. The ortho/para ratio for T$_r$=5\,K
is 0.1 and the column density for the para species is ten times higher than for the ortho species. However,
for T$_r$=10 K, the column density of the para species decreases by two orders of magnitude and
the ortho/para ratio increases to a value of 1.
We re-examined their column densities using the data shown in their Figure \ref{fig_hdccn}. Considering,
as we have for TMC-1, that the ortho and para spin symmetry states are two different species, we
derive for this cloud $N$(o-H$_2$CCN)=$4 \times 10^{12}$ cm$^{-2}$ and $N$(p-H$_2$CCN)=$5 \times 10^{11}$ cm$^{-2}$.
Hence, the ortho/para ratio in L1544 is $\sim$8, that is, very similar to that found in TMC-1. Similar
results are obtained for T$_r$=10 K.

\begin{table}
\tiny
\caption{H$_2$CCN and HDCCN column densities in TMC-1.}
\label{tab_CD}
\centering
\begin{tabular}{{l|c|c|c|c|c|c}}
\hline
\hline
Parameter/T$_{rot}$                         &    5 K &   6 K &   7 K & 8 K & 9 K & 10 K\\
\hline
N($o$-H$_2$CCN)$^a$                         & 131   &  121  & 118   & 120 & 124 & 128\\
N($p$-H$_2$CCN)$^a$                         &  17   &   17  &  17   &  17 &  18 &  19\\
N(H$_2$CCN)$^a$                             & 148   &  138  & 135   & 137 & 142 & 147\\
N($o$-H$_2$CCN)/N($p$-H$_2$CCN)             & 7.7   &  7.1  & 7.0   & 7.1 & 6.9 & 6.7\\
N(HDCCN)$^a$                                & 7.4   &  7.5  & 8.0   & 8.8 & 9.6 & 10.5\\
N(H$_2$CCN)/N(HDCCN)                        &20.0   & 18.4  & 16.9  & 15.6& 14.8& 14.0\\
\hline
\end{tabular}
\tablefoot{
        \tablefoottext{a}{Column densities are in units of 10$^{11}$ cm$^{-2}$. The uncertainty on the
     derived column densities is $\sim$10\% based on the adopted calibration accuracy
     (see Sec.~\ref{observations}). The adopted line width for all features of H$_2$CCN and HDCCN is
     0.6 km\,s$^{-1}$.}
}
\end{table}
\normalsize

\begin{table}
\tiny
\caption{Deuteration enhancement in TMC-1 for selected molecules compared to our gas-phase chemical model.}
\label{tab_H/D}
\centering
\begin{tabular}{{l|c|c|c|c}}
\hline
\hline
                         &         & model $^a$ & model $^a$ &  model$^a$ \\
Molecule                 &   TMC-1 &  N/O = 0.5 & N/O = 1 &  N/O=1.5  \\
\hline
H$_2$CCN/HDCCN           &   20$^b$  &  17.6   &  16.2   &   15.2  \\
HC$_3$N/DC$_3$N$^c$      &   62      &  78.8   &  75.3   &   73.0  \\
HNCCC/DNCCC$^c$          &   43      &  34.8   &  35.9   &   37.2  \\
HCCNC/DCCNC$^c$          &   30      &  71.7   &  68.3   &   65.9  \\
HCNCC/DCNCC$^d$          &           &  21.2   &  20.9   &   20.6  \\
HC$_5$N/DC$_5$N$^c$      &   82      &  27.8   &  28.3   &   29.4  \\
CH$_3$CCH/CH$_3$CCD      &   49      &  204.6  &  210.   &  213.3  \\
CH$_3$CCH/CH$_2$DCCH     &   10$^e$  &   68.2  &   70.1  &   71.1  \\
CH$_3$CN/CH$_2$DCN       &   11$^e$  &   10.2  &   9.9   &   9.7   \\
c-C$_3$H$_2$/c-C$_3$HD   &   27$^b$  &   45.5  &   45.1  &   45.1  \\
H$_2$C$_4$/HDC$_4$       &  83$^b$   &  114.4  &  113.6  &   114.4 \\
C$_4$H/C$_4$D            &  118      &  104.3  &  107.1  &   109.5 \\

\hline
\end{tabular}
\tablefoot{
    \tablefoottext{a}{See text for the assumed physical conditions. [O/H] = $8 \times 10^{-6}$, D/H = $1.5 \times 10^{-5}$,
           and C/O = 0.75.}
        \tablefoottext{b}{The elemental H/D abundance ratio is affected by the
     presence of two identical nuclei of H.}
        \tablefoottext{c}{These H/D abundance ratios are from  \citet{Cernicharo2020e}.}
        \tablefoottext{d}{This isomer of HC$_3$N has not yet been detected in space.}
        \tablefoottext{e}{The elemental H/D abundance ratio is affected by the
     presence of three identical nuclei of H.}
}
\end{table}
\normalsize

\subsection{Chemical models} \label{chemical models}

After the surprise raised by the detection of DCN in the Orion cloud  by
\cite{Jefferts74}, it was very quickly suggested  that the observed deuterium
enrichment is the result of chemical fractionation and not of an exceptional source of deuterium through nucleosynthetic processes.
The first determinant step is provided by the deuterium exchange in the H$_3^+$ + HD $\rightleftarrows$ H$_2$D$^+$ + H$_2$ reaction. This reaction proceeds essentially in the forward direction in the cold (T $\sim$ 10K) interstellar environments, as the endothermicity of the reverse reaction is about 232 K, which  results from the difference in zero point energies (ZPE) of the products and reactants. This crucial point was first mentioned by \cite{Watson1976} and extended to the D + H$_3^+$ $\rightleftarrows$  H + H$_2$D$^+$ reaction by \cite{dalgarno84}, which proceeds essentially in the forward direction as the endothermicity of the reverse reaction is then 632K. The  H$_2$D$^+$ molecular ion subsequently transfers the deuteron to various
stable molecules, such as CO, N$_2$, H$_2$O, NH$_3$, producing DCO$^+$, N$_2$D$^+$,
H$_2$DO$^+$, NH$_3$D$^+$, which disassociatively recombine to produce, for example, HDO and NH$_2$D.
This scenario has been further extended to a few other stable molecular ions such as
CH$_3^+$ and C$_2$H$_3^+$, invoking exchange reactions with HD as in CH$_3^+$ + HD $\rightleftarrows$ CH$_2$D$^+$ + H$_2$ and
C$_2$H$_3^+$ + HD $\rightleftarrows$ C$_2$H$_2$D$^+$ + H$_2$ \citep{herbst87,gerlich02}.
The exothermicity achieved in the forward reactions are somewhat larger than
that involving H$_2$D$^+$ \citep{herbst87,roueff13,nyman19}, allowing these
exchange reactions to occur in somewhat warmer environments such as clumps in the
Orion Bar \citep{roueff07,parise09}. These exchange reactions can even proceed
further toward doubly deuterated molecular ions like D$_2$H$^+$ and CHD$_2^+$, and up to
fully deuterated ions like D$_3^+$, CD$_3^+$, and C$_2$D$_3^+$ in environments with CO depleted
\citep{walmsley04}, where the ion-molecule reactions with CO, H$_2$O, and NH$_3$ become negligible in comparison to the reactions with HD (see, e.g., \citealt{roberts03,roueff05}).
Additional deuterium enhancements could result from  the evaporation of the icy mantles surrounding
dust grain as a remnant of cold grain chemistry.
Gas-phase chemical processes are reasonably successful to interpret the observations,
as reported recently by \cite{bacmann20} and \cite{Melosso2020} in their study of the nitrogen hydrides NHD and ND$_2$.

We introduced the deuteration of the carbon chains C$_n$H, C$_n$H$_2$, and nitriles in our gas-phase chemical network, which has been used to study the influence of the C/O ratio on the chemistry of nitriles by \cite{legal19}. That study benefitted from the recent
updates reported by \cite{loison14,loison17} for nitrogen and carbon-chain chemistry,
including the CH$_3$CN isomers.  We did not introduce an additional deuterium exchange reaction in the network because the various exchange reactions with HD that  we considered
(HCNH$^+$, HC$_3$NH$^+$, CH$_3$CNH$^+$) are all hindered by the presence of a barrier of
about 56 kcal/mol (see Appendix \ref{ab_initio}). Then,
deuterium fractionation results directly from the enrichment of H$_2$D$^+$ and, to a lesser
extent, from CH$_2$D$^+$ and C$_2$H$_2$D$^+$. In the case of HDCCN, the main route to deuteration is
provided by the reaction H$_2$CCN + H$_2$D$^+$ $\rightarrow$ CH$_2$DCN$^+$ + H$_2$, followed by CH$_2$DCN$^+$ + $e^-$ $\rightarrow$ HDCCN + H.
The various dissociative channels of that reaction are reported in \cite{loison14}
for the hydrogenated ions. We assume, as in our previous studies on nitrogen deuteration
\citep{roueff05}, that in absence of experimental information the total rate coefficient of dissociative recombination of the deuterated ion is identical to that of the hydrogenated
molecular ion. We hypothesize that hydrogen ejection is twice as efficient as deuterium ejection, as found experimentally for HDO$^+$ \citep{jensen00}.
Our final gas-phase chemical network includes 288 chemical species and more than
7000 chemical reactions\footnote{The initial network \citep{legal19} comprised 190
chemical species and about 3500 chemical reactions.}. More details on our gas-phase chemical
network will be described in a later publication. We ran steady-state models that are appropriate
for the TMC-1 conditions, that is, n(H$_2$) = 10$^4$ cm$^{-3}$, T=10 K, $\zeta$ = $1.3 \times 10^{-17}$
s$^{-1}$, and deplete oxygen to get the observed fractional abundance of CO relative
to molecular hydrogen of $\sim 1.2 \times 10^{-5}$ \citep{cernicharo87}. The results are shown
in Table \ref{tab_H/D} and compared to the observational ratios. The deuterium ratio of H$_2$CCN
is found to be remarkably stable, independent of the assumed N/O ratio and  close to the observational
value. The other deuterium ratios are within about a factor of two of the observed values, with a couple of exceptions, and are not
very sensitive to the N/O ratio. The HCCNC isomer is found by the observations to be more fractionated than
theoretically predicted, which may indicate specific chemical rearrangements in the concerned
chemical reactions. The main disagreement is obtained for deuterated methyl acetylene, where some additional deuterium enrichment
mechanism could be missing. Further studies are desirable.

\section{Conclusions}

 We detected, and fully characterized spectroscopically, the
singly deuterated isotopolog of H$_2$CCN from astronomical and laboratory measurements. The observed deuterium ratio for this and other species is well accounted for by a standard gas-phase model, in which deuteration is transferred through the H$_2$D$^+$ molecular ion, followed by dissociative recombination.

\begin{acknowledgements}

The Spanish authors thank Ministerio de Ciencia e Innovaci\'on for funding
support through project AYA2016-75066-C2-1-P, PID2019-106235GB-I00 and
PID2019-107115GB-C21 / AEI / 10.13039/501100011033. We also thank ERC for funding through grant
ERC-2013-Syg-610256-NANOCOSMOS. MA thanks Ministerio de Ciencia e Innovaci\'on
for grant RyC-2014-16277. Y. Endo thanks Ministry of Science
and Technology of Taiwan through grant MOST108-2113-M-009-25. Some of the kinetic data used for
the chemical model have been taken from the online database KIDA \citep{wakelam15},
http://kida.obs.u-bordeaux1.fr.

\end{acknowledgements}

\normalsize

\begin{appendix}

\section{Quantum chemical calculation}
\label{ab_initio}

Quantum chemical calculations were carried out to estimate the molecular parameters of HDCCN.
Very precise values for the rotational parameters of HDCCN can be obtained using experimetal/theoretical
ratios derived for H$_2$CCN species.
This is the most common method to predict the expected experimental rotational constants for an isotopic species
of a given molecule
when the rotational constants for its parent species are known. The geometry optimization calculations were done
using
a coupled cluster with single, double, and perturbative triple excitation (CCSD(T); \citealt{Raghavachari1989})
methods with all electrons
(valence and core) correlated and Dunning's correlation-consistent basis sets with polarized core-valence
correlation quadruple-$\zeta$
(cc-pCVQZ; \citealt{Woon1995}). This calculations were carried out using the Molpro 2018.1 program
\citep{Werner2018}. Table \ref{abini} shows the results for these calculations.
In addition to the rotational constants, we calculated other parameters necessary to interpret the rotational
spectrum of HDCCN. In this manner,
the three spin-rotation coupling constants ($\varepsilon_{aa}$, $\varepsilon_{bb}$ and $\varepsilon_{cc}$),
the magnetic hyperfine constants ($a_F$, $T_{aa}$ and $T_{bb}$), and the nuclear electric quadrupole constants for the deuterium nucleus
($\chi_{aa}$ and $\chi_{bb}$) were estimated following the same procedure used for the
rotational constants. However, in this case the optimized molecular structure at CCSD(T)/cc-pCVQZ was used.
These calculations were carried out via the Becke, three-parameter, Lee–Yang–Parr (B3LYP) hybrid density functional
and Dunning's correlation-consistent basis sets with correlation-consistent
polarized valence quadruple-$\zeta$ (cc-pVQZ; \citealt{Dunning1989}). The results for these calculations
carried out using the Gaussian16 \citep{Frisch2016} program package are shown in Table \ref{abini}.

\begin{table}
\tiny
\caption{Experimental, theoretical, and scaled values for the spectroscopic parameters of H$_2$CCN and HDCCN(all in MHz).}
\label{abini}
\centering
\begin{tabular}{{lcccc}}
\hline
\hline
&\multicolumn{2}{c}{H$_2$CCN}&\multicolumn{2}{c}{HDCCN} \\
\hline
Parameter & Calc.\tablefootmark{a} & Exp.\tablefootmark{b} & Calc.\tablefootmark{a} & Scaled\tablefootmark{c}\\
\hline
$A$                                 &   287050.9       & 284981.         &     197241.4799     &  195819.1       \\
$B$                 &  10252.6       &  10246.7658       &      9608.3332      &     9602.3      \\
$C$                 &  9899.0        &  9876.029         &      9162.0187      &     9141.2      \\
\hline
$\varepsilon_{aa}$  &  -776.794      &  -661.537                 &      -533.755            &      -454.558     \\
$\varepsilon_{bb}$  &   -27.537      &  -24.1205         &        -25.807      &     -22.605     \\
$\varepsilon_{cc}$  &     1.059      &  -2.0345          &          0.980      &       -1.883    \\
\hline
$a_F$ (H)           &    -54.9478    &  -59.63           &                     &                 \\
$T_{aa}$ (H)        &    -18.2057    &  -15.9006         &                     &                 \\
$T_{bb}$ (H)        &    17.7382     &  14.6             &                     &                 \\
\hline
$a_F$ (D)           &                &                   &      -8.434         &     -9.153      \\
$T_{aa}$ (D)        &                &                   &      -2.219         &     -2.156      \\
$T_{bb}$ (D)        &                &                   &       2.147         &      1.972      \\
\hline
$\chi_{aa}$ (N)     &       -4.195   &      -4.7705      &                     &                 \\
$\chi_{bb}$ (N)     &       1.8327   &      2.0341       &                     &                 \\
\hline
$\chi_{aa}$ (D)     &                &                   &       -0.0111       &     -0.0097     \\
$\chi_{bb}$ (D)     &                &                   &        0.0165       &      0.1048     \\
\hline
\end{tabular}
\tablefoot{\\
\tablefoottext{a}{Rotational constants calculated at CCSD(T)/cc-pCVQZ level of theory and the rest of parameters at
at B3LYP/cc-pVQZ level of theory.}
\tablefoottext{b}{\cite{Saito1997} and \cite{Ozeki2004}.}
\tablefoottext{c}{Scaled by the ratio Exp/Calc. of the corresponding parameter for the H$_2$CCN species}.
}
\normalsize
\end{table}

In addition to the structural optimizations, quantum chemical calculations were carried out to examine the potential energy surface for the deuteration reactions of HCNH$^+$, HC$_3$NH$^+$, CH$_3$CNH$^+$ with HD. All the possible stationary points were fully optimized using the coupled cluster singles and doubles correlated method (CCSD; \citealt{Cizek1969}) along with Dunning’s correlation-consistent polarized valence triple-$\zeta$ (cc-pVTZ; \citealt{Dunning1989}) basis set. The starting point, with the cationic species (HCNH$^+$, HC$_3$NH$^+$ or CH$_3$CNH$^+$) and HD separated from each other, was assumed to be energy zero. The reactions proceed initially by the formation of a pre-reactive complex (HCNH$^+$-HD, HC$_3$NH$^+$-HD, and  CH$_3$CNH$^+$-HD), which have energies 2.23, 2.09, and 2.85 kcal/mol lower than the separated reactants for HCNH$^+$, HC$_3$NH$^+$ and CH$_3$CNH$^+$, respectively. The barrier heights of the transition states between the pre-reactive complexes and the deuterated species (HCND$^+$, HC$_3$ND$^+$, CH$_3$CND$^+$) are 56.42, 51.87, and 52.76 kcal/mol, respectively.

\section{Laboratory transition frequencies for HDCCN}
\label{lab_freq}
The laboratory measurements of HDCCN described in Sect.\,\ref{labo} permitted us to measure 132 hyperfine components in the $N_{K_{a,}K_{c}}$=2$_{0,2}$-1$_{0,1}$
and 1$_{0,1}$-0$_{0,0}$ rotational transitions. The observed frequencies, quantum
number assignments, and calculated line intensities are given in Table \ref{lab_freq}.
We uploaded a file with predictions up to $N$=16 to the CDS.

\onecolumn
\begin{longtable}{cccccccccccccccccc}
\caption[]{Laboratory observed transition frequencies for HDCCN.
\label{tab_lab_freq}}\\
\hline
\hline
 $N'$ & $K'_a$ & $K'_c$ & $J'$ & $F'_1$ & $F'_2$ & $F'$ & $N''$ & $K''_a$ & $K''_c$ & $J''$ & $F''_1$ & $F''_2$ & $F''$ & $\nu_{obs}$  &  $\nu_{calc}$  &   Obs-Calc & Int.$^a$ \\
      &        &        &      &        &        &      &      &        &        &      &        &        &      &   (MHz)      &   (MHz)        &   (MHz)    &      \\
\hline
\endfirsthead
\caption{continued.}\\
\hline
\hline
 $N'$ & $K'_a$ & $K'_c$ & $J'$ & $F'_1$ & $F'_2$ & $F'$ & $N''$ & $K''_a$ & $K''_c$ & $J''$ & $F''_1$ & $F''_2$ & $F''$ & $\nu_{obs}$  &  $\nu_{calc}$  &   Obs-Calc & Int.$^a$ \\
      &        &        &      &        &        &      &      &        &        &      &        &        &      &   (MHz)      &   (MHz)        &   (MHz)    &      \\
\hline
\endhead
\hline
\endfoot
\hline
\endlastfoot
\hline
1 & 0 & 1 & 0.5 & 1.5 & 1 & 2 & 0 & 0 & 0 & 0.5 & 1.5 & 1 & 2 &  18708.037  & 18708.035 &~  0.002   &  0.065    \\
1 & 0 & 1 & 1.5 & 1.5 & 2 & 1 & 0 & 0 & 0 & 0.5 & 1.5 & 2 & 1 &  18729.487  & 18729.484 &~  0.003   &  0.068    \\
1 & 0 & 1 & 1.5 & 1.5 & 2 & 3 & 0 & 0 & 0 & 0.5 & 1.5 & 2 & 3 &  18733.323  & 18733.327 &~  -0.004 &   0.460   \\
1 & 0 & 1 & 1.5 & 1.5 & 2 & 2 & 0 & 0 & 0 & 0.5 & 1.5 & 2 & 2 &  18734.829  & 18734.828 &~  0.001   &  0.464    \\
1 & 0 & 1 & 1.5 & 0.5 & 1 & 1 & 0 & 0 & 0 & 0.5 & 0.5 & 1 & 1 &  18735.210  & 18735.212 &~  -0.002 &   0.093   \\
1 & 0 & 1 & 1.5 & 0.5 & 1 & 1 & 0 & 0 & 0 & 0.5 & 0.5 & 1 & 0 &  18735.956  & 18735.958 &~  -0.002 &   0.106   \\
1 & 0 & 1 & 1.5 & 1.5 & 1 & 2 & 0 & 0 & 0 & 0.5 & 1.5 & 2 & 3 &  18736.776  & 18736.776 &~  0.000   &  0.089    \\
1 & 0 & 1 & 1.5 & 0.5 & 1 & 2 & 0 & 0 & 0 & 0.5 & 0.5 & 1 & 2 &  18737.107  & 18737.113 &~  -0.006 &   0.307   \\
1 & 0 & 1 & 1.5 & 1.5 & 1 & 0 & 0 & 0 & 0 & 0.5 & 0.5 & 1 & 1 &  18738.806  & 18738.809 &~  -0.003 &   0.163   \\
1 & 0 & 1 & 1.5 & 1.5 & 1 & 2 & 0 & 0 & 0 & 0.5 & 0.5 & 1 & 1 &  18739.142  & 18739.144 &~  -0.002 &   0.220   \\
1 & 0 & 1 & 0.5 & 1.5 & 2 & 1 & 0 & 0 & 0 & 0.5 & 1.5 & 2 & 1 &  18740.523  & 18740.520 &~  0.003   &  0.440    \\
1 & 0 & 1 & 0.5 & 1.5 & 2 & 2 & 0 & 0 & 0 & 0.5 & 1.5 & 2 & 2 &  18740.684  & 18740.681 &~  0.003   &  0.669    \\
1 & 0 & 1 & 1.5 & 0.5 & 1 & 2 & 0 & 0 & 0 & 0.5 & 0.5 & 0 & 1 &  18741.148  & 18741.154 &~  -0.006 &   1.347   \\
1 & 0 & 1 & 1.5 & 0.5 & 1 & 1 & 0 & 0 & 0 & 0.5 & 0.5 & 1 & 2 &  18741.432  & 18741.432 &~  0.000   &  0.144    \\
1 & 0 & 1 & 0.5 & 1.5 & 2 & 3 & 0 & 0 & 0 & 0.5 & 1.5 & 2 & 3 &  18741.500  & 18741.497 &~  0.003   &  1.866    \\
1 & 0 & 1 & 1.5 & 1.5 & 2 & 3 & 0 & 0 & 0 & 0.5 & 0.5 & 1 & 2 &  18741.910  & 18741.914 &~  -0.004 &   1.865   \\
1 & 0 & 1 & 1.5 & 0.5 & 0 & 1 & 0 & 0 & 0 & 0.5 & 0.5 & 1 & 1 &  18742.375  & 18742.376 &~  -0.001 &   0.564   \\
1 & 0 & 1 & 1.5 & 2.5 & 3 & 2 & 0 & 0 & 0 & 0.5 & 1.5 & 2 & 1 &  18743.128  & 18743.128 &~  0.000   &  1.509    \\
1 & 0 & 1 & 1.5 & 2.5 & 3 & 3 & 0 & 0 & 0 & 0.5 & 1.5 & 2 & 2 &  18743.343  & 18743.343 &~  0.000   &  2.320    \\
1 & 0 & 1 & 1.5 & 2.5 & 3 & 4 & 0 & 0 & 0 & 0.5 & 1.5 & 2 & 3 &  18743.468  & 18743.469 &~  -0.001 &   3.000   \\
1 & 0 & 1 & 1.5 & 1.5 & 2 & 1 & 0 & 0 & 0 & 0.5 & 0.5 & 1 & 0 &  18744.633  & 18744.631 &~  0.002   &  0.699    \\
1 & 0 & 1 & 1.5 & 1.5 & 2 & 2 & 0 & 0 & 0 & 0.5 & 0.5 & 1 & 1 &  18744.711  & 18744.713 &~  -0.002 &   0.888   \\
1 & 0 & 1 & 0.5 & 1.5 & 2 & 1 & 0 & 0 & 0 & 0.5 & 1.5 & 2 & 2 &  18745.042  & 18745.037 &~  0.005   &  0.363    \\
1 & 0 & 1 & 1.5 & 2.5 & 2 & 1 & 0 & 0 & 0 & 0.5 & 1.5 & 1 & 0 &  18745.118  & 18745.116 &~  0.002   &  0.261    \\
1 & 0 & 1 & 1.5 & 1.5 & 1 & 2 & 0 & 0 & 0 & 0.5 & 0.5 & 1 & 2 &  18745.363  & 18745.364 &~  -0.001 &   1.079   \\
1 & 0 & 1 & 1.5 & 0.5 & 1 & 1 & 0 & 0 & 0 & 0.5 & 0.5 & 0 & 1 &  18745.471  & 18745.474 &~  -0.003 &   0.645   \\
1 & 0 & 1 & 1.5 & 0.5 & 1 & 0 & 0 & 0 & 0 & 0.5 & 0.5 & 1 & 1 &  18745.751  & 18745.760 &~  -0.009 &   0.129   \\
1 & 0 & 1 & 1.5 & 2.5 & 2 & 1 & 0 & 0 & 0 & 0.5 & 1.5 & 1 & 1 &  18745.785  & 18745.783 &~  0.002   &  0.500    \\
1 & 0 & 1 & 1.5 & 2.5 & 2 & 2 & 0 & 0 & 0 & 0.5 & 1.5 & 1 & 1 &  18746.086  & 18746.083 &~  0.003   &  1.017    \\
1 & 0 & 1 & 0.5 & 0.5 & 1 & 1 & 0 & 0 & 0 & 0.5 & 1.5 & 1 & 0 &  18746.812  & 18746.811 &~  0.001   &  0.480    \\
1 & 0 & 1 & 1.5 & 2.5 & 2 & 1 & 0 & 0 & 0 & 0.5 & 1.5 & 1 & 2 &  18747.164  & 18747.160 &~  0.004   &  0.209    \\
1 & 0 & 1 & 1.5 & 2.5 & 2 & 3 & 0 & 0 & 0 & 0.5 & 1.5 & 1 & 2 &  18747.320  & 18747.318 &~  0.002   &  2.316    \\
1 & 0 & 1 & 0.5 & 0.5 & 1 & 0 & 0 & 0 & 0 & 0.5 & 1.5 & 1 & 1 &  18747.339  & 18747.335 &~  0.004   &  0.308    \\
1 & 0 & 1 & 1.5 & 2.5 & 2 & 2 & 0 & 0 & 0 & 0.5 & 1.5 & 1 & 2 &  18747.462  & 18747.460 &~  0.002   &  0.624    \\
1 & 0 & 1 & 1.5 & 1.5 & 1 & 1 & 0 & 0 & 0 & 0.5 & 0.5 & 1 & 1 &  18747.734  & 18747.733 &~  0.001   &  0.228    \\
1 & 0 & 1 & 0.5 & 1.5 & 2 & 2 & 0 & 0 & 0 & 0.5 & 1.5 & 2 & 3 &  18748.202  & 18748.199 &~  0.003   &  0.554    \\
1 & 0 & 1 & 1.5 & 0.5 & 0 & 1 & 0 & 0 & 0 & 0.5 & 0.5 & 1 & 2 &  18748.595  & 18748.596 &~  -0.001 &   0.124   \\
1 & 0 & 1 & 0.5 & 0.5 & 1 & 1 & 0 & 0 & 0 & 0.5 & 1.5 & 1 & 2 &  18748.856  & 18748.855 &~  0.001   &  0.455    \\
1 & 0 & 1 & 1.5 & 1.5 & 1 & 0 & 0 & 0 & 0 & 0.5 & 0.5 & 0 & 1 &  18749.069  & 18749.070 &~  -0.001 &   0.164   \\
1 & 0 & 1 & 1.5 & 1.5 & 1 & 2 & 0 & 0 & 0 & 0.5 & 0.5 & 0 & 1 &  18749.404  & 18749.405 &~  -0.001 &   0.225   \\
1 & 0 & 1 & 0.5 & 0.5 & 1 & 2 & 0 & 0 & 0 & 0.5 & 1.5 & 1 & 1 &  18749.583  & 18749.582 &~  0.001   &  0.623    \\
1 & 0 & 1 & 0.5 & 1.5 & 2 & 3 & 0 & 0 & 0 & 0.5 & 0.5 & 1 & 2 &  18750.085  & 18750.085 &~  0.000   &  0.454    \\
1 & 0 & 1 & 1.5 & 1.5 & 2 & 1 & 0 & 0 & 0 & 0.5 & 0.5 & 1 & 2 &  18750.111  & 18750.106 &~  0.005   &  0.170    \\
1 & 0 & 1 & 0.5 & 1.5 & 2 & 2 & 0 & 0 & 0 & 0.5 & 0.5 & 1 & 1 &  18750.568  & 18750.567 &~  0.001   &  0.401    \\
1 & 0 & 1 & 1.5 & 1.5 & 2 & 2 & 0 & 0 & 0 & 0.5 & 0.5 & 1 & 2 &  18750.923  & 18750.933 &~  -0.010 &   0.191   \\
1 & 0 & 1 & 0.5 & 0.5 & 1 & 2 & 0 & 0 & 0 & 0.5 & 1.5 & 1 & 2 &  18750.959  & 18750.959 &~  0.000   &  0.935    \\
1 & 0 & 1 & 1.5 & 0.5 & 0 & 1 & 0 & 0 & 0 & 0.5 & 0.5 & 0 & 1 &  18752.636  & 18752.637 &~  -0.001 &   0.224   \\
1 & 0 & 1 & 1.5 & 1.5 & 1 & 1 & 0 & 0 & 0 & 0.5 & 0.5 & 1 & 2 &  18753.952  & 18753.954 &~  -0.002 &   0.473   \\
1 & 0 & 1 & 0.5 & 0.5 & 0 & 1 & 0 & 0 & 0 & 0.5 & 1.5 & 1 & 0 &  18753.986  & 18753.983 &~  0.003   &  0.234    \\
1 & 0 & 1 & 0.5 & 0.5 & 0 & 1 & 0 & 0 & 0 & 0.5 & 1.5 & 1 & 1 &  18754.648  & 18754.650 &~  -0.002 &   0.462   \\
1 & 0 & 1 & 0.5 & 1.5 & 2 & 1 & 0 & 0 & 0 & 0.5 & 0.5 & 1 & 1 &  18754.925  & 18754.923 &~  0.002   &  0.093    \\
1 & 0 & 1 & 1.5 & 1.5 & 2 & 2 & 0 & 0 & 0 & 0.5 & 0.5 & 0 & 1 &  18754.971  & 18754.975 &~  -0.004 &   0.083   \\
1 & 0 & 1 & 1.5 & 2.5 & 3 & 2 & 0 & 0 & 0 & 0.5 & 1.5 & 2 & 3 &  18755.163  & 18755.163 &~  0.000   &  0.085    \\
1 & 0 & 1 & 0.5 & 1.5 & 2 & 1 & 0 & 0 & 0 & 0.5 & 0.5 & 1 & 0 &  18755.670  & 18755.668 &~  0.002   &  0.090    \\
1 & 0 & 1 & 0.5 & 0.5 & 0 & 1 & 0 & 0 & 0 & 0.5 & 1.5 & 1 & 2 &  18756.024  & 18756.027 &~  -0.003 &   0.264   \\
1 & 0 & 1 & 1.5 & 0.5 & 1 & 0 & 0 & 0 & 0 & 0.5 & 0.5 & 0 & 1 &  18756.024  & 18756.021 &~  0.003   &   0.152   \\
1 & 0 & 1 & 0.5 & 1.5 & 1 & 2 & 0 & 0 & 0 & 0.5 & 1.5 & 2 & 1 &  18756.391  & 18756.390 &~  0.001   &   0.116   \\
1 & 0 & 1 & 1.5 & 1.5 & 1 & 1 & 0 & 0 & 0 & 0.5 & 0.5 & 0 & 1 &  18757.992  & 18757.995 &~  -0.003 &         0.071  \\
1 & 0 & 1 & 0.5 & 1.5 & 1 & 2 & 0 & 0 & 0 & 0.5 & 1.5 & 2 & 2 &  18760.907  & 18760.906 &~  0.001   &   0.427   \\
1 & 0 & 1 & 0.5 & 1.5 & 1 & 1 & 0 & 0 & 0 & 0.5 & 1.5 & 2 & 1 &  18762.014  & 18762.013 &~  0.001   &   0.320   \\
1 & 0 & 1 & 0.5 & 1.5 & 1 & 0 & 0 & 0 & 0 & 0.5 & 1.5 & 2 & 1 &  18764.270  & 18764.269 &~  0.001   &   0.258   \\
1 & 0 & 1 & 0.5 & 1.5 & 1 & 1 & 0 & 0 & 0 & 0.5 & 1.5 & 2 & 2 &  18766.530  & 18766.530 &~  0.000   &   0.489   \\
1 & 0 & 1 & 0.5 & 1.5 & 1 & 2 & 0 & 0 & 0 & 0.5 & 1.5 & 2 & 3 &  18768.425  & 18768.424 &~  0.001   &   0.828   \\
1 & 0 & 1 & 0.5 & 1.5 & 1 & 2 & 0 & 0 & 0 & 0.5 & 0.5 & 1 & 1 &  18770.792  & 18770.792 &~  0.000   &   0.141   \\
1 & 0 & 1 & 0.5 & 1.5 & 1 & 2 & 0 & 0 & 0 & 0.5 & 0.5 & 1 & 2 &  18777.011  & 18777.012 &~  -0.001 &         0.072  \\
1 & 0 & 1 & 0.5 & 1.5 & 1 & 1 & 0 & 0 & 0 & 0.5 & 0.5 & 1 & 2 &  18782.635  & 18782.635 &~  0.000   &   0.062   \\
2 & 0 & 2 & 2.5 & 2.5 & 3 & 3 & 1 & 0 & 1 & 1.5 & 2.5 & 3 & 3 &  37473.508  & 37473.508 &~  0.000   &   0.087   \\
2 & 0 & 2 & 2.5 & 2.5 & 3 & 4 & 1 & 0 & 1 & 1.5 & 2.5 & 3 & 4 &  37477.170  & 37477.173 &~  -0.003 &         0.298  \\
2 & 0 & 2 & 2.5 & 2.5 & 2 & 2 & 1 & 0 & 1 & 1.5 & 2.5 & 3 & 2 &  37479.963  & 37479.958 &~  0.005   &   0.189   \\
2 & 0 & 2 & 2.5 & 3.5 & 4 & 3 & 1 & 0 & 1 & 1.5 & 2.5 & 3 & 2 &  37481.282  & 37481.280 &~  0.002   &   0.055   \\
2 & 0 & 2 & 2.5 & 2.5 & 2 & 3 & 1 & 0 & 1 & 1.5 & 2.5 & 3 & 3 &  37481.398  & 37481.393 &~  0.005   &   0.234   \\
2 & 0 & 2 & 2.5 & 1.5 & 2 & 3 & 1 & 0 & 1 & 1.5 & 1.5 & 2 & 3 &  37481.934  & 37481.940 &~  -0.006 &         0.241  \\
2 & 0 & 2 & 2.5 & 1.5 & 2 & 2 & 1 & 0 & 1 & 1.5 & 1.5 & 1 & 2 &  37482.690  & 37482.687 &~  0.003   &   0.172   \\
2 & 0 & 2 & 2.5 & 3.5 & 4 & 3 & 1 & 0 & 1 & 1.5 & 2.5 & 3 & 3 &  37485.589  & 37485.583 &~  0.006   &   0.210   \\
2 & 0 & 2 & 2.5 & 1.5 & 2 & 1 & 1 & 0 & 1 & 1.5 & 1.5 & 1 & 2 &  37486.615  & 37486.607 &~  0.008   &   0.054   \\
2 & 0 & 2 & 2.5 & 1.5 & 2 & 2 & 1 & 0 & 1 & 1.5 & 0.5 & 1 & 1 &  37486.620  & 37486.619 &~  0.001   &   1.285   \\
2 & 0 & 2 & 2.5 & 2.5 & 3 & 2 & 1 & 0 & 1 & 1.5 & 1.5 & 2 & 1 &  37486.636  & 37486.630 &~  0.006   &   0.224   \\
2 & 0 & 2 & 2.5 & 1.5 & 2 & 3 & 1 & 0 & 1 & 1.5 & 0.5 & 1 & 2 &  37486.748  & 37486.742 &~  0.006   &   2.547   \\
2 & 0 & 2 & 2.5 & 1.5 & 2 & 1 & 1 & 0 & 1 & 1.5 & 1.5 & 1 & 0 &  37486.940  & 37486.942 &~  -0.002 &         0.503  \\
2 & 0 & 2 & 2.5 & 2.5 & 3 & 4 & 1 & 0 & 1 & 1.5 & 1.5 & 2 & 3 &  37487.310  & 37487.315 &~  -0.005 &         3.286  \\
2 & 0 & 2 & 2.5 & 1.5 & 1 & 2 & 1 & 0 & 1 & 1.5 & 1.5 & 2 & 2 &  37487.342  & 37487.343 &~  -0.001 &         0.209  \\
2 & 0 & 2 & 1.5 & 2.5 & 3 & 3 & 1 & 0 & 1 & 1.5 & 2.5 & 3 & 2 &  37487.579  & 37487.574 &~  0.005   &   2.372   \\
2 & 0 & 2 & 2.5 & 2.5 & 3 & 3 & 1 & 0 & 1 & 1.5 & 1.5 & 1 & 2 &  37487.597  & 37487.592 &~  0.005   &   1.833   \\
2 & 0 & 2 & 2.5 & 3.5 & 4 & 5 & 1 & 0 & 1 & 1.5 & 2.5 & 3 & 4 &  37487.598  & 37487.598 &~  0.000   &   4.400   \\
2 & 0 & 2 & 2.5 & 1.5 & 1 & 2 & 1 & 0 & 1 & 1.5 & 1.5 & 2 & 1 &  37488.174  & 37488.170 &~  0.004   &   1.211   \\
2 & 0 & 2 & 1.5 & 2.5 & 2 & 2 & 1 & 0 & 1 & 0.5 & 1.5 & 1 & 2 &  37488.564  & 37488.556 &~  0.008   &   0.325   \\
2 & 0 & 2 & 2.5 & 2.5 & 2 & 3 & 1 & 0 & 1 & 1.5 & 1.5 & 2 & 2 &  37489.902  & 37489.907 &~  -0.005 &         2.209  \\
2 & 0 & 2 & 1.5 & 2.5 & 2 & 3 & 1 & 0 & 1 & 0.5 & 1.5 & 1 & 2 &  37490.414  & 37490.414 &~  0.000   &   1.400   \\
2 & 0 & 2 & 2.5 & 1.5 & 2 & 1 & 1 & 0 & 1 & 1.5 & 0.5 & 1 & 1 &  37490.539  & 37490.539 &~  0.000   &   0.348   \\
2 & 0 & 2 & 2.5 & 1.5 & 2 & 2 & 1 & 0 & 1 & 1.5 & 0.5 & 1 & 2 &  37490.936  & 37490.938 &~  -0.002 &         0.444  \\
2 & 0 & 2 & 2.5 & 2.5 & 3 & 3 & 1 & 0 & 1 & 1.5 & 1.5 & 2 & 3 &  37491.042  & 37491.041 &~  0.001   &   0.622   \\
2 & 0 & 2 & 2.5 & 1.5 & 1 & 1 & 1 & 0 & 1 & 1.5 & 1.5 & 2 & 1 &  37491.506  & 37491.500 &~  0.006   &   0.059   \\
2 & 0 & 2 & 2.5 & 3.5 & 3 & 2 & 1 & 0 & 1 & 1.5 & 2.5 & 2 & 2 &  37492.147  & 37492.152 &~  -0.005 &         0.473  \\
2 & 0 & 2 & 2.5 & 3.5 & 3 & 2 & 1 & 0 & 1 & 1.5 & 2.5 & 2 & 1 &  37492.449  & 37492.452 &~  -0.003 &         1.394  \\
2 & 0 & 2 & 1.5 & 0.5 & 1 & 1 & 1 & 0 & 1 & 0.5 & 0.5 & 0 & 1 &  37492.480  & 37492.487 &~  -0.007 &         0.462  \\
2 & 0 & 2 & 2.5 & 2.5 & 2 & 2 & 1 & 0 & 1 & 1.5 & 1.5 & 2 & 2 &  37492.774  & 37492.776 &~  -0.002 &         0.170  \\
2 & 0 & 2 & 2.5 & 3.5 & 4 & 3 & 1 & 0 & 1 & 1.5 & 2.5 & 3 & 4 &  37492.978  & 37492.974 &~  0.004   &   0.114   \\
2 & 0 & 2 & 2.5 & 2.5 & 2 & 1 & 1 & 0 & 1 & 1.5 & 0.5 & 1 & 0 &  37493.001  & 37493.007 &~  -0.006 &         0.132  \\
2 & 0 & 2 & 2.5 & 3.5 & 3 & 3 & 1 & 0 & 1 & 1.5 & 2.5 & 2 & 2 &  37493.031  & 37493.034 &~  -0.003 &         2.285  \\
2 & 0 & 2 & 2.5 & 3.5 & 3 & 3 & 1 & 0 & 1 & 1.5 & 2.5 & 2 & 3 &  37493.173  & 37493.176 &~  -0.003 &         0.468  \\
2 & 0 & 2 & 2.5 & 2.5 & 2 & 2 & 1 & 0 & 1 & 1.5 & 1.5 & 2 & 1 &  37493.607  & 37493.603 &~  0.004   &   0.156   \\
2 & 0 & 2 & 1.5 & 2.5 & 3 & 2 & 1 & 0 & 1 & 0.5 & 1.5 & 2 & 2 &  37493.802  & 37493.804 &~  -0.002 &         0.785  \\
2 & 0 & 2 & 2.5 & 1.5 & 1 & 0 & 1 & 0 & 1 & 1.5 & 1.5 & 2 & 1 &  37493.916  & 37493.909 &~  0.007   &   0.065   \\
2 & 0 & 2 & 1.5 & 2.5 & 3 & 4 & 1 & 0 & 1 & 1.5 & 2.5 & 2 & 3 &  37494.060  & 37494.067 &~  -0.007 &         3.562  \\
2 & 0 & 2 & 1.5 & 1.5 & 2 & 1 & 1 & 0 & 1 & 1.5 & 2.5 & 2 & 2 &  37494.093  & 37494.087 &~  0.006   &   0.102   \\
2 & 0 & 2 & 1.5 & 1.5 & 2 & 1 & 1 & 0 & 1 & 1.5 & 2.5 & 2 & 1 &  37494.386  & 37494.387 &~  -0.001 &         0.443  \\
2 & 0 & 2 & 1.5 & 1.5 & 1 & 0 & 1 & 0 & 1 & 0.5 & 1.5 & 1 & 1 &  37494.952  & 37494.962 &~  -0.010 &         0.229  \\
2 & 0 & 2 & 1.5 & 1.5 & 2 & 2 & 1 & 0 & 1 & 1.5 & 2.5 & 2 & 2 &  37495.683  & 37495.686 &~  -0.003 &         0.261  \\
2 & 0 & 2 & 1.5 & 1.5 & 2 & 2 & 1 & 0 & 1 & 1.5 & 2.5 & 2 & 3 &  37495.826  & 37495.828 &~  -0.002 &         0.073  \\
2 & 0 & 2 & 2.5 & 2.5 & 2 & 1 & 1 & 0 & 1 & 1.5 & 0.5 & 0 & 1 &  37496.386  & 37496.391 &~  -0.005 &         0.229  \\
2 & 0 & 2 & 1.5 & 1.5 & 2 & 3 & 1 & 0 & 1 & 1.5 & 2.5 & 2 & 2 &  37497.641  & 37497.645 &~  -0.004 &         0.116  \\
2 & 0 & 2 & 1.5 & 1.5 & 2 & 3 & 1 & 0 & 1 & 1.5 & 2.5 & 2 & 3 &  37497.787  & 37497.787 &~  0.000   &   0.441   \\
2 & 0 & 2 & 2.5 & 2.5 & 2 & 2 & 1 & 0 & 1 & 1.5 & 1.5 & 1 & 2 &  37498.348  & 37498.345 &~  0.003   &   0.273   \\
2 & 0 & 2 & 2.5 & 2.5 & 2 & 3 & 1 & 0 & 1 & 1.5 & 1.5 & 2 & 3 &  37498.931  & 37498.926 &~  0.005   &   0.200   \\
2 & 0 & 2 & 1.5 & 1.5 & 1 & 1 & 1 & 0 & 1 & 0.5 & 1.5 & 1 & 2 &  37499.444  & 37499.439 &~  0.005   &   0.444   \\
2 & 0 & 2 & 1.5 & 2.5 & 3 & 2 & 1 & 0 & 1 & 1.5 & 1.5 & 2 & 2 &  37499.651  & 37499.657 &~  -0.006 &         0.052  \\
2 & 0 & 2 & 2.5 & 1.5 & 1 & 1 & 1 & 0 & 1 & 1.5 & 0.5 & 1 & 1 &  37500.173  & 37500.174 &~  -0.001 &         0.116  \\
2 & 0 & 2 & 1.5 & 2.5 & 3 & 2 & 1 & 0 & 1 & 0.5 & 1.5 & 2 & 3 &  37500.509  & 37500.506 &~  0.003   &   0.145   \\
2 & 0 & 2 & 2.5 & 1.5 & 1 & 2 & 1 & 0 & 1 & 1.5 & 0.5 & 1 & 2 &  37501.169  & 37501.163 &~  0.006   &   0.140   \\
2 & 0 & 2 & 1.5 & 1.5 & 1 & 2 & 1 & 0 & 1 & 0.5 & 1.5 & 1 & 2 &  37501.191  & 37501.185 &~  0.006   &   0.771   \\
2 & 0 & 2 & 1.5 & 2.5 & 3 & 3 & 1 & 0 & 1 & 0.5 & 1.5 & 2 & 3 &  37501.238  & 37501.240 &~  -0.002 &         0.231  \\
2 & 0 & 2 & 2.5 & 2.5 & 2 & 2 & 1 & 0 & 1 & 1.5 & 1.5 & 2 & 3 &  37501.794  & 37501.795 &~  -0.001 &         0.074  \\
2 & 0 & 2 & 1.5 & 2.5 & 2 & 2 & 1 & 0 & 1 & 1.5 & 2.5 & 3 & 2 &  37501.804  & 37501.817 &~  -0.013 &         0.272  \\
2 & 0 & 2 & 1.5 & 2.5 & 2 & 3 & 1 & 0 & 1 & 1.5 & 2.5 & 3 & 3 &  37507.971  & 37507.978 &~  -0.007 &         0.316  \\
2 & 0 & 2 & 1.5 & 2.5 & 2 & 1 & 1 & 0 & 1 & 0.5 & 1.5 & 2 & 2 &  37508.394  & 37508.394 &~  0.000   &   0.254   \\
2 & 0 & 2 & 1.5 & 0.5 & 0 & 1 & 1 & 0 & 1 & 0.5 & 1.5 & 1 & 2 &  37508.720  & 37508.720 &~  0.000   &   0.209   \\
2 & 0 & 2 & 1.5 & 2.5 & 2 & 2 & 1 & 0 & 1 & 0.5 & 1.5 & 2 & 2 &  37508.780  & 37508.781 &~  -0.001 &         0.069  \\
2 & 0 & 2 & 1.5 & 2.5 & 2 & 3 & 1 & 0 & 1 & 0.5 & 1.5 & 2 & 2 &  37510.641  & 37510.639 &~  0.002   &   0.404   \\
2 & 0 & 2 & 1.5 & 1.5 & 1 & 2 & 1 & 0 & 1 & 1.5 & 2.5 & 3 & 3 &  37518.740  & 37518.749 &~  -0.009 &         0.055  \\
\hline
\end{longtable}
\tablefoot{$^a$Calculated line strength factor.\\}
\twocolumn

\section{Observed deuteration in other molecular species}
\label{deuteration}
In order to compare the derived deuteration enhancement in H$_2$CCN with that of
other species, we searched in our data for the lines of
the main and singly deuterated isotopologs of CH$_3$CN, CH$_3$CCH,
c-C$_3$H$_2$, C$_4$H, and H$_2$C$_4$. The line parameters for the observed
rotational transitions of these species are
given in Table \ref{tab_deuterated}.
For most of these species, only one or two low-$J$ lines
were observed in the line survey. We  assume a rotational
temperature of 10 K for these species. We would like to stress that for most of the
species analyzed the variation in column density between T$_r$=5-10 K
is rather small as was the case for HDCCN (see Sect. \ref{CD_HDCCN}).
We observe several
rotational transitions for C$_6$H and a rotational diagram was performed.
The derived XH/XD abundances are given in Table \ref{tab_H/D}. We now comment on each molecule.

In our survey CH$_3$CCH only has a rotational transition for each symmetry
species $A$ and $E$. The data have previously been shown to compare their $A$/$E$
abundance ratio with that of CH$_3$CO$^+$ \citep{Cernicharo2020c}.
The derived line parameters are given in Table \ref{tab_deuterated}. The total
column density ($A$+$E$) of CH$_3$CCH for T$_r$=10 K is (1.3$\pm$0.1)$\times$10$^{14}$ cm$^{-2}$.
Adopting a T$_r$=5 K, we derive practically the same column density.
Taking into account the low dipole moment of this species, we could expect that its
lowest energy levels are thermalized at the kinetic temperature of the cloud.
Two single-deuterated species are observed. We obtain a
column density for T$_r$=10 of (2.8$\pm$0.3)$\times$10$^{12}$ cm$^{-2}$ from the line parameters
of CH$_3$CCD given in Table \ref{tab_deuterated}.
The H/D abundance ratio for this deuterated isotologue is $\sim$48.
The other single-deuterated substitute, CH$_2$DCCH, is an asymmetric rotor
and several lines are observed in our line survey. We derive a column density
for T$_r$=10 K of (1.4$\pm$0.1)$\times$10$^{13}$ cm$^{-2}$ from the derived line
parameters (see Table \ref{tab_deuterated}). This value
decreases by 15\% if a rotational temperature of 5 K is adopted. The
H/D abundance ratio for this deuterated isotopolog is $\sim$10. To
compare with the value derived for CH$_3$CCD, we have to take into account a factor of 3 in the abundance ratio coming from the three identical
possible substitutions of H by D ina the methyl group. It seems that
deuteration in the methyl group is slightly favored (H/D=30) with respect
to deuteration in the acetylenic group of the molecule (H/D=48). The three
$^{13}$C isotopologs were detected. Using the integrated lines
intensity, we derive an averaged $^{12}$C/$^{13}$C ratio of 75$\pm$10. This
abundance ratio is in very good agreement with the value derived by
\citet{Cernicharo2020e} from the $^{13}$C isotopologs of HC$_5$N.

We also show the data previously in the same context
as for CH$_3$CCH \citep{Cernicharo2020c} for CH$_3$CN. The line parameters for all
hyperfine components of this molecule are given
in Table \ref{tab_deuterated}. We derive N($A$-CH$_3$CN)=(2.5$\pm$0.2)$\times$10$^{12}$
cm$^{-2}$ and N($E$-CH$_3$CN)=(2.2$\pm$0.2)$\times$10$^{12}$ cm$^{-2}$ (T$_r$=10 K).
For CH$_2$DCN, we detect three rotational transitions exhibiting
hyperfine splitting. To the best of our knowledge, this hyperfine structure
has only been observed in the early years of microwave spectroscopy \citep{Thomas1955}.
The reported frequencies, often blends of several hyperfine components, are shifted by
50-100 kHz with respect to the observed frequencies in TMC-1. From these new rest
frequencies we derive $\chi_{aa}$=-4.194$\pm$0.012 MHz. We find that $\chi_{bb}$ was
not necessary to reproduce the observed transitions within our frequency
accuracy (10 kHz). The column density derived for this deuterated species is (4.5$\pm$0.6)$\times$10$^{11}$
cm$^{-2}$ and the N(CH$_3$CN)/N(CH$_2$DCN) abundance ratio is $\sim$11, that is, very similar to the
value found for the N(CH$_3$CCH)/N(CH$_2$DCCH) abundance ratio.

We observed the 3$_{2,1}$-3$_{1,2}$ $ortho$ line at 44104.777 MHz for c-C$_3$H$_2$.
This line appears in absorption against the cosmic background radiation.
This
line was not detected in the line survey of TMC-1 performed with the Nobeyama telescope \citep{Kaifu2004}.
In addition, we observed in three $para$ lines in our data (see Table \ref{tab_deuterated}). Of these
three lines, only the strongest one was reported by \citealt{Kaifu2004}. One of the $para$ lines (4$_{40}$-3$_{31}$)
also appears in our data
in absorption against the cosmic background radiation. Hence, it is not possible to adopt a uniform rotational
temperature for the lines of both $ortho$ and $para$ species. Nevertheless, collisional rates are available
for the system c-C$_3$H$_2$/He
\citep{Avery1989}, which allows us to perform excitation models using the large velocity gradient approximation.
For the typical volume density of TMC-1 of 4$\times$10$^4$ cm$^{-3}$ and T$_K$=10 K, the three $para$ lines are
very well reproduced, including the one in absorption that has a column density for this species of
N($p$-c-C$_3$H$_2$)=(3.2$\pm$0.3)$\times$10$^{13}$ cm$^{-2}$.
The absorption produced by the $ortho$ line shows a very strong dependence on the density for
a large range of column densities. The
line is predicted in absorption for n(H$_2$)$\leq$3$\times$10$^4$ cm$^{-3}$ and
N($o$-c-C$_3$H$_2$)$\geq$5$\times$10$^{13}$ cm$^{-2}$.
The line is predicted
to be in emission for larger densities and/or lower column densities. We found that
n(H$_2$)=(2-3)$\times$10$^4$ cm$^{-3}$ and N(o-c-C$_3$H$_2$)=(9$\pm$1.0)$\times$10$^{13}$ cm$^{-2}$ reproduce the observed
absorption of this species. Hence, the total column density of c-C$_3$H$_2$ is (1.2$\pm$0.1)$\times$10$^{14}$ cm$^{-2}$, that is,
very similar to that of CH$_3$CCH. The derived $ortho/para$ ratio is 2.8$\pm$0.6.
The column density and rotational temperature of c-C$_3$HD can be derived from the four observed lines
to be (4.5$\pm$0.5)$\times$10$^{12}$ cm$^{-2}$ and T$_r$=6$\pm$1 K, respectively. Therefore, the
H/D abundance ratio in this species is 27$\pm$5.

Six lines of H$_2$C$_4$  and two of HDC$_4$were found in our survey. The line parameters
are given in Table \ref{tab_deuterated}. We explored the column densities and rotational
temperatures that reproduce at the best the observed line intensities. We find that for T$_r$=10 K
N($o$-H$_2$C$_4$)=(3.0$\pm$0.3)$\times$10$^{12}$ cm$^{-2}$ and
N($o$-H$_2$C$_4$)=(1.0$\pm$0.1)$\times$10$^{12}$ cm$^{-2}$. Similar results are obtained for T$_r$=7 K.
For T$_r$=5 K, the column densities have to be increased by a 25\%.
The $ortho/para$ abundance ratio is 3$\pm$0.3 for this species. We
obtain N(HDC$_4$)=(4.8$\pm$0.5)$\times$10$^{10}$ cm$^{-2}$ for T$_r$=10 K. Consequently, the H/D abundance ratio for
H$_2$C$_4$ is 83$\pm$15.

For C$_4$H, we modeled the emission for all the lines in our survey, including the weak hyperfine
lines that are typically a factor 30-50 weaker than the strongest components. The best fit is
obtained for T$_r$=7 K and N(C$_4$H)=(1.3$\pm$0.1)$\times$10$^{14}$ cm$^{-2}$, although a model
with the same column density and T$_r$=10 K provides practically the same result.
The column density increases by a 20\% for T$_r$=5 K. Adopting T$_r$=7 K for C$_4$D, we derive
N(C$_4$D)=(1.1$\pm$0.1)$\times$10$^{11}$ cm$^{-2}$. Hence, the H/D abundance ratio for C$_4$H is
118$\pm$20.

\onecolumn
\begin{longtable}{llcrccr}
\caption[]{Observed line parameters for selected molecules and their deuterated isotopologs in TMC-1.
\label{tab_deuterated}}\\
\hline
\hline
\multicolumn{1}{c}{Molecule}&\multicolumn{1}{c}{Quantum Numbers$^a$}&\multicolumn{1}{c}{$\nu_{rest}^b$ }&\multicolumn{1}{c}{$\int$T$_A^*$dv $^c$}&\multicolumn{1}{c}{v$_{LSR}$$^d$}&\multicolumn{1}{c}{$\Delta$v$^e$}&\multicolumn{1}{c}{T$_A^*$$^f$}  \\
& &\multicolumn{1}{c}{(MHz)}&\multicolumn{1}{c}{(mK km\,s$^{-1}$)}&\multicolumn{1}{c}{(km\,s$^{-1}$)}&\multicolumn{1}{c}{(km\,s$^{-1}$)}&\multicolumn{1}{c}{(mK)}  \\
\hline
\endfirsthead
\caption{continued.}\\
\hline
\hline
\multicolumn{1}{c}{Molecule}&\multicolumn{1}{c}{Quantum Numbers$^a$}&\multicolumn{1}{c}{$\nu_{rest}^b$ }&\multicolumn{1}{c}{$\int$T$_A^*$dv $^c$}&\multicolumn{1}{c}{v$_{LSR}$$^d$}&\multicolumn{1}{c}{$\Delta$v$^e$}&\multicolumn{1}{c}{T$_A^*$$^f$}  \\
& &\multicolumn{1}{c}{(MHz)}&\multicolumn{1}{c}{(mK km\,s$^{-1}$)}&\multicolumn{1}{c}{(km\,s$^{-1}$)}&\multicolumn{1}{c}{(km\,s$^{-1}$)}&\multicolumn{1}{c}{(mK)}  \\
\hline
\endhead
\hline
\endfoot
\hline
\endlastfoot
\hline
CH$_3$CCH  & 2 0-1 0         &34183.4139$\pm$0.0001&  338.0$\pm$0.5& 5.79$\pm$0.01& 0.76$\pm$0.01& 274.4$\pm$0.8\\
CH$_3$CCH  & 2 1-1 1         &34182.7603$\pm$0.0001&  262.5$\pm$0.5& 5.79$\pm$0.01& 0.75$\pm$0.01& 209.3$\pm$0.8\\
\hline
CH$_3$CCD  & 2 0-1 0         &31152.6032$\pm$0.0007&    5.4$\pm$1.5& 5.87$\pm$0.05& 0.83$\pm$0.12&   6.1$\pm$0.8\\
CH$_3$CCD  & 2 1-1 1         &31152.0321$\pm$0.0007&    4.2$\pm$1.5& 5.73$\pm$0.09& 0.99$\pm$0.17&   4.0$\pm$0.8\\
CH$_3$CCD  & 3 0-2 0         &46728.7668$\pm$0.0010&    9.2$\pm$1.1& 5.79$\pm$0.02& 0.58$\pm$0.04&  14.8$\pm$0.8\\
CH$_3$CCD  & 3 1-2 1         &46727.9102$\pm$0.0010&    7.6$\pm$1.0& 5.82$\pm$0.02& 0.53$\pm$0.04&  13.5$\pm$0.8\\
\hline
CH$_2$DCCH & 2 1 2-1 1 1     &32231.5022$\pm$0.0019&    7.4$\pm$0.9& 5.80$\pm$0.02& 0.86$\pm$0.05&   8.1$\pm$0.5\\
CH$_2$DCCH & 2 0 2-1 0 1     &32362.1261$\pm$0.0020&   20.0$\pm$0.6& 5.77$\pm$0.01& 0.73$\pm$0.01&  25.7$\pm$0.5\\
CH$_2$DCCH & 2 1 1-1 1 0     &32491.9172$\pm$0.0019&    8.6$\pm$0.9& 5.80$\pm$0.02& 0.89$\pm$0.05&   9.1$\pm$0.5\\
CH$_2$DCCH & 3 1 3-2 1 2     &48347.0251$\pm$0.0028&   18.0$\pm$1.5& 5.79$\pm$0.01& 0.62$\pm$0.03&  27.4$\pm$1.0\\
CH$_2$DCCH & 3 2 2-2 2 1     &48540.0261$\pm$0.0027&    3.0$\pm$0.6& 5.95$\pm$0.06& 0.60$\pm$0.12&   4.8$\pm$1.1\\
CH$_2$DCCH & 3 2 1-2 2 0     &48540.4725$\pm$0.0027&               &              &              &   $\leq$3.3  \\
CH$_2$DCCH & 3 0 3-2 0 2     &48542.7476$\pm$0.0029&   33.8$\pm$2.3& 5.79$\pm$0.01& 0.60$\pm$0.03&  53.2$\pm$1.1\\ 
CH$_2$DCCH & 3 1 2-2 1 1     &48737.6403$\pm$0.0029&   15.5$\pm$1.5& 5.80$\pm$0.01& 0.59$\pm$0.03&  24.7$\pm$1.1\\
\hline
$^{13}$CH$_3$CCH& 2 0-1 0    &33252.9070$\pm$0.0002&    2.3$\pm$0.6& 5.77$\pm$0.04& 0.65$\pm$0.10&   3.3$\pm$0.4\\
$^{13}$CH$_3$CCH& 2 1-1 1    &33252.2861$\pm$0.0002&    2.6$\pm$0.6& 5.93$\pm$0.05& 0.76$\pm$0.10&   3.2$\pm$0.4\\
$^{13}$CH$_3$CCH& 3 0-2 0    &49879.1918$\pm$0.0002&    5.2$\pm$2.1& 5.83$\pm$0.06& 0.50$\pm$0.12&   9.6$\pm$2.1\\
$^{13}$CH$_3$CCH& 3 1-2 1    &49878.2606$\pm$0.0002&    6.2$\pm$2.1& 5.93$\pm$0.04& 0.49$\pm$0.08&  11.9$\pm$2.1\\
\hline
CH$_3$$^{13}$CCH& 2 0-1 0    &34169.2361$\pm$0.0001&    3.2$\pm$0.6& 5.75$\pm$0.03& 0.76$\pm$0.07&   3.9$\pm$0.4\\
CH$_3$$^{13}$CCH& 2 1-1 1    &34168.5850$\pm$0.0002&    2.4$\pm$0.8& 5.84$\pm$0.06& 0.95$\pm$0.15&   2.3$\pm$0.4\\
\hline
CH$_3$C$^{13}$CH& 2 0-1 0    &33160.8989$\pm$0.0002&    3.1$\pm$0.7& 5.81$\pm$0.04& 0.82$\pm$0.09&   3.6$\pm$0.4\\
CH$_3$C$^{13}$CH& 2 1-1 1    &33160.2761$\pm$0.0002&    1.8$\pm$0.7& 5.82$\pm$0.04& 0.54$\pm$0.14&   3.2$\pm$0.4\\
CH$_3$C$^{13}$CH& 3 0-2 0    &49741.1817$\pm$0.0003&    2.7$\pm$1.4& 5.82$\pm$0.09& 0.52$\pm$0.07&   4.9$\pm$1.9\\
CH$_3$C$^{13}$CH& 3 1-2 1    &49740.2489$\pm$0.0003&    3.7$\pm$1.7& 5.91$\pm$0.07& 0.53$\pm$0.09&   6.5$\pm$1.9\\
\hline
CH$_3$CN        & 2 1-1 1 2-1&36793.7090$\pm$0.0003&   39.0$\pm$0.9& 5.74$\pm$0.01& 0.68$\pm$0.01&  53.6$\pm$0.4\\ 
CH$_3$CN        & 2 0-1 0 2-2&36794.2044$\pm$0.0003&   19.9$\pm$0.6& 5.75$\pm$0.01& 0.68$\pm$0.01&  27.6$\pm$0.4\\ 
CH$_3$CN        & 2 1-1 1 2-2&36794.3402$\pm$0.0002&   20.0$\pm$0.8& 5.61$\pm$0.05& 0.84$\pm$0.02&  22.5$\pm$0.4\\ 
CH$_3$CN        & 2 0-1 0 1-0&36794.4173$\pm$0.0002&   19.6$\pm$0.6& 5.69$\pm$0.05& 0.60$\pm$0.01&  30.6$\pm$0.4\\ 
CH$_3$CN        & 2 1-1 1 1-1&36794.7618$\pm$0.0002&   12.3$\pm$0.7& 5.74$\pm$0.01& 0.66$\pm$0.02&  17.6$\pm$0.4\\ 
CH$_3$CN        & 2 1-1 1 3-2&36795.0234$\pm$0.0001&   69.5$\pm$1.3& 5.73$\pm$0.01& 0.67$\pm$0.01&  98.2$\pm$0.4\\ 
CH$_3$CN        & 2 0-1 0 2-1&36795.4747$\pm$0.0001&   51.6$\pm$1.1& 5.79$\pm$0.01& 0.63$\pm$0.01&  77.2$\pm$0.4\\ 
CH$_3$CN        & 2 0-1 0 3-2&36795.5669$\pm$0.0001&  113.6$\pm$1.9& 5.77$\pm$0.01& 0.72$\pm$0.01& 147.8$\pm$0.4\\ 
CH$_3$CN        & 2 1-1 1 1-0&36796.3466$\pm$0.0005&   21.8$\pm$0.9& 5.74$\pm$0.01& 0.76$\pm$0.02&  27.1$\pm$0.4\\ 
CH$_3$CN        & 2 0-1 0 1-1&36797.5829$\pm$0.0006&   20.9$\pm$0.9& 5.74$\pm$0.01& 0.69$\pm$0.02&  28.4$\pm$0.4\\ 
\hline
CH$_2$DCN & 2$_{12}$-1$_{11}$ 2-1 & 34583.1020$\pm$0.0100&  2.6$\pm$0.7& 5.83$\pm$0.00& 0.84$\pm$0.14&  2.9$\pm$0.4\\
CH$_2$DCN & 2$_{12}$-1$_{11}$ 2-2 & 34583.1020$\pm$0.0100&             &              &              &             \\
CH$_2$DCN & 2$_{12}$-1$_{11}$ 3-2 & 34584.4110$\pm$0.0100&  2.1$\pm$0.0& 5.83$\pm$0.00& 0.61$\pm$0.10&  3.2$\pm$0.4\\
CH$_2$DCN & 2$_{02}$-1$_{01}$ 2-2 & 34733.9580$\pm$0.0100&  1.6$\pm$0.0& 5.83$\pm$0.00& 0.51$\pm$0.08&  2.9$\pm$0.4\\
CH$_2$DCN & 2$_{02}$-1$_{01}$ 1-0 & 34734.1710$\pm$0.0100&  1.3$\pm$0.0& 5.83$\pm$0.00& 0.35$\pm$0.08&  3.8$\pm$0.4\\
CH$_2$DCN & 2$_{02}$-1$_{01}$ 2-1 & 34735.2280$\pm$0.0100&  5.8$\pm$0.0& 5.83$\pm$0.00& 0.75$\pm$0.07&  7.2$\pm$0.4\\
CH$_2$DCN & 2$_{02}$-1$_{01}$ 3-2 & 34735.3190$\pm$0.0100& 10.6$\pm$0.0& 5.83$\pm$0.00& 0.72$\pm$0.04& 13.9$\pm$0.4\\
CH$_2$DCN & 2$_{02}$-1$_{01}$ 1-1 & 34737.3130$\pm$0.0100&  1.7$\pm$0.0& 5.83$\pm$0.00& 0.70$\pm$0.09&  2.4$\pm$0.4\\
CH$_2$DCN & 2$_{11}$-1$_{10}$ 2-1 & 34884.3920$\pm$0.0100&  1.6$\pm$0.0& 5.83$\pm$0.00& 0.88$\pm$0.18&  1.8$\pm$0.3\\
CH$_2$DCN & 2$_{11}$-1$_{10}$ 1-1 & 34884.3920$\pm$0.0100&             &              &              &             \\
CH$_2$DCN & 2$_{11}$-1$_{10}$ 3-2 & 34885.7090$\pm$0.0100&  2.2$\pm$0.0& 5.83$\pm$0.00& 0.62$\pm$0.08&  3.3$\pm$0.3\\
CH$_2$DCN & 2$_{11}$-1$_{10}$ 2-2 & 34885.7090$\pm$0.0100&             &              &              &             \\
\hline
c-C$_3$H$_2$&3$_{21}$-3$_{12}$&44104.7766$\pm$0.0014&  -29.5$\pm$   & 5.66$\pm$0.01& 0.53$\pm$0.01& -52.4$\pm$0.7\\
c-C$_3$H$_2$&4$_{40}$-3$_{31}$&35360.9275$\pm$0.0018&   -2.0$\pm$0.0& 5.88$\pm$0.05& 0.62$\pm$0.08&  -3.0$\pm$0.4\\
c-C$_3$H$_2$&4$_{31}$-4$_{22}$&42231.2517$\pm$0.0012&    3.5$\pm$0.7& 5.73$\pm$0.03& 0.63$\pm$0.07&   5.1$\pm$0.5\\
c-C$_3$H$_2$&2$_{11}$-2$_{02}$&46755.6096$\pm$0.0013&  514.1$\pm$8.8& 5.75$\pm$0.02& 0.62$\pm$0.01& 780.7$\pm$0.7\\
\hline
c-C$_3$HD  &3$_{21}$-3$_{12}$ &35600.4822$\pm$0.0030&    4.8$\pm$0.6& 5.85$\pm$0.02& 0.69$\pm$0.05&   6.5$\pm$0.4\\
c-C$_3$HD  &2$_{11}$-2$_{02}$ &38224.4450$\pm$0.0030&   54.1$\pm$1.0& 5.83$\pm$0.01& 0.76$\pm$0.01&  66.8$\pm$0.4\\
c-C$_3$HD  &3$_{13}$-2$_{20}$ &39605.9580$\pm$0.0080&               &              &              &     $\leq$1.5\\
c-C$_3$HD  &1$_{01}$-0$_{00}$ &42064.1467$\pm$0.0020&   29.8$\pm$1.1& 5.72$\pm$0.05& 0.79$\pm$0.02&  35.3$\pm$0.4\\ 
c-C$_3$HD  &4$_{31}$-4$_{22}$ &44828.8655$\pm$0.0040&               &              &              &     $\leq$2.4\\
c-C$_3$HD  &3$_{30}$-3$_{21}$ &47788.7950$\pm$0.0060&               &              &              &     $\leq$4.5\\
c-C$_3$HD  &1$_{11}$-0$_{00}$ &49615.8560$\pm$0.0026&  142.0$\pm$3.7& 5.78$\pm$0.01& 0.65$\pm$0.01& 205.9$\pm$2.2\\
\hline
H$_2$C$_4$ &4$_{14}$-3$_{13}$ &35577.0079$\pm$0.0015&  143.0$\pm$0.0& 5.71$\pm$0.01& 0.68$\pm$0.01& 197.8$\pm$0.4\\
H$_2$C$_4$ &4$_{04}$-3$_{03}$ &35727.3789$\pm$0.0011&   92.3$\pm$0.0& 5.56$\pm$0.01& 0.65$\pm$0.01& 132.7$\pm$0.5\\
H$_2$C$_4$ &4$_{13}$-3$_{12}$ &35875.7746$\pm$0.0016&  143.2$\pm$0.0& 5.70$\pm$0.01& 0.65$\pm$0.01& 206.2$\pm$0.4\\
H$_2$C$_4$ &5$_{15}$-4$_{14}$ &44471.1375$\pm$0.0018&  123.2$\pm$0.0& 5.72$\pm$0.01& 0.48$\pm$0.01& 243.3$\pm$0.7\\
H$_2$C$_4$ &5$_{05}$-4$_{04}$ &35727.3789$\pm$0.0011&   75.2$\pm$0.0& 5.57$\pm$0.01& 0.48$\pm$0.01& 146.3$\pm$0.8\\
H$_2$C$_4$ &5$_{14}$-4$_{13}$ &44844.5902$\pm$0.0020&  123.2$\pm$0.0& 5.71$\pm$0.01& 0.53$\pm$0.01& 217.0$\pm$0.8\\
\hline
HDC$_4$    & 4$_{04}$-3$_{03}$&33902.1130$\pm$0.0030&   4.6$\pm$0.0 & 5.72$\pm$0.02& 0.73$\pm$0.05&   6.0$\pm$0.4\\
HDC$_4$    & 5$_{05}$-4$_{04}$&42377.2757$\pm$0.0080&   4.5$\pm$0.0 & 5.73$\pm$0.03& 0.56$\pm$0.07&   7.6$\pm$0.6\\
\hline
CCCCH & 4-3  9/2-7/2 4-3      &38049.6160$\pm$0.0010& 550.6$\pm$9.4& 5.78$\pm$0.01& 0.60$\pm$0.01&  862.1$\pm$0.4\\
CCCCH & 4-3  9/2-7/2 5-4      &38049.6910$\pm$0.0010& 641.9$\pm$9.8& 5.78$\pm$0.01& 0.60$\pm$0.01& 1005.0$\pm$0.4\\
CCCCH & 4-3  9/2-7/2 4-4      &38059.4310$\pm$0.0020&  24.6$\pm$0.6& 5.74$\pm$0.01& 0.65$\pm$0.01&   35.8$\pm$0.4\\
CCCCH & 4-3  7/2-5/2 3-3      &38078.9300$\pm$0.0020&  25.1$\pm$0.6& 5.74$\pm$0.01& 0.65$\pm$0.01&   36.0$\pm$0.4\\
CCCCH & 4-3  7/2-5/2 4-3 + 3-2&38088.4600$\pm$0.0010& 913.3$\pm$9.9& 5.64$\pm$0.01& 0.77$\pm$0.01& 1120.8$\pm$0.5\\
CCCCH & 4-3  7/2-7/2 3-3      &38212.6370$\pm$0.0040&  21.6$\pm$0.7& 5.75$\pm$0.01& 0.58$\pm$0.01&   34.8$\pm$0.5\\
CCCCH & 4-3  7/2-7/2 3-4      &38231.9620$\pm$0.0040&  23.9$\pm$0.7& 5.74$\pm$0.01& 0.60$\pm$0.01&   37.5$\pm$0.5\\
CCCCH & 5-4 11/2-9/2 5-4 + 6-5&47566.7920$\pm$0.0020& 953.1$\pm$9.9& 5.64$\pm$0.01& 0.69$\pm$0.01& 1297.1$\pm$1.2\\
CCCCH & 5-4 11/2-9/2 5-5      &47576.5100$\pm$0.0020&  14.0$\pm$1.5& 5.72$\pm$0.01& 0.51$\pm$0.03&   25.8$\pm$1.2\\
CCCCH & 5-4  9/2-7/2 4-4      &47595.9910$\pm$0.0020&  15.1$\pm$1.5& 5.69$\pm$0.01& 0.52$\pm$0.03&   27.2$\pm$1.2\\
CCCCH & 5-4  9/2-7/2 5-4 + 4-3&47605.4950$\pm$0.0020& 761.4$\pm$9.9& 5.78$\pm$0.01& 0.58$\pm$0.01& 1224.6$\pm$1.2\\
CCCCH & 5-4  9/2-9/2 4-4      &47768.5220$\pm$0.0040&  15.5$\pm$1.4& 5.76$\pm$0.01& 0.55$\pm$0.02&   26.5$\pm$1.4\\
CCCCH & 5-4  9/2-9/2 5-5      &47787.7600$\pm$0.0040&  15.0$\pm$1.0& 5.68$\pm$0.01& 0.49$\pm$0.01&   28.8$\pm$1.4\\
\hline
CCCCD       & 4-3  9/2-7/2    &35313.3131$\pm$0.0020&  17.3$\pm$0.0& 5.72$\pm$0.01& 0.89$\pm$0.02&   18.3$\pm$0.4\\
CCCCD       & 4-3  7/2-5/2    &35349.4296$\pm$0.0020&  13.6$\pm$0.0& 5.78$\pm$0.01& 0.89$\pm$0.03&   14.4$\pm$0.4\\
CCCCD       & 5-4 11/2-9/2    &44146.0436$\pm$0.0102&  18.5$\pm$0.0& 5.73$\pm$0.01& 0.63$\pm$0.02&   27.6$\pm$0.8\\
CCCCD       & 5-4  9/2-7/2    &44182.1087$\pm$0.0102&  15.4$\pm$0.0& 5.77$\pm$0.01& 0.65$\pm$0.02&   22.4$\pm$0.8\\
\hline
\end{longtable}
\tablefoot{
    \tablefoottext{a}{Quantum numbers are $J K$ for symmetric rotors, $J_{K_{a},K_{c}}$
    for asymmetric rotors. If
    the molecule contains an atom of nitrogen, then the additional quantum number corresponds to $F$, the total
    angular momentum of the molecule.}\\
        \tablefoottext{b}{Rest frequencies from MADEX \citep{Cernicharo2012} and/or the CMDS \citep{Muller2005}.
    If the uncertainty on the velocity is zero, then the observed frequency is given fixing the
    velocity of the source to 5.83 km\,s$^{-1}$.}\\
        \tablefoottext{c}{Integrated line intensity in mK km\,s$^{-1}$.}\\
        \tablefoottext{d}{Local standard of rest velocity of the line in mK km\,s$^{-1}$. If the uncertainty
     is equal to zero, then the velocity is fixed to 5.83 km\,s$^{-1}$.}\\
        \tablefoottext{e}{Line width at half intensity derived by fitting a Gaussian line profile to the observed
     transitions (in km\,s$^{-1}$).}\\
        \tablefoottext{f}{Antenna temperature and 1$\sigma$ uncertainty in mK. For upper
     limits we provide a 3$\sigma$ value.}\\
}
\end{appendix}

\end{document}